\input{psfig.sty}
\documentclass[11pt,a4paper]{article}
\usepackage{jcappub}

\title{A unified solution to the small scale problems of the $\Lambda$CDM model}
\author[a,b]{A. Del Popolo,}
\author[c]{J. A. S. Lima,}
\author[d]{J\'ulio C. Fabris,}
\author[d]{Davi C. Rodrigues}

\affiliation[a]{Dipartimento di Fisica e Astronomia, University Of Catania, \\
Viale Andrea Doria 6, 95125 Catania, Italy \\}
\affiliation[b]{International Institute of Physics, Universidade Federal do Rio Grande do Norte, Av. Odilon Gomes de Lima, 1722 - Capim Macio - 59078-400, Natal-RN, Brazil}
\affiliation[c]{Departamento de Astronomia, Universidade de S\~ao Paulo, Rua do Mat\~ao 1226, 05508-900, S\~ao Paulo, SP,
Brazil}
\affiliation[d]{Departamento de F\'isica, Universidade Federal do Esp\'irito Santo, Av. Fernando Ferrari 514, Vit\'oria - ES, Brazil}

\emailAdd{adelpopolo@oact.inaf.it;limajas@astro.iag.usp.br; fabris@pq.cnpq.br; davi.rodrigues@cosmo-ufes.org}
\abstract{We study, by means of the model proposed in Del Popolo (2009), the effect of baryon physics on the small scale problems of the CDM model. We show that, using this model, the cusp/core problem, the missing satellite problem (MSP), the Too Big to Fail (TBTF) problem, and the angular momentum catastrophe can be reconciled  with observations. Concerning the cusp/core problem, the interaction  among dark matter (DM) and baryonic clumps of 1\% the mass of the halo,  through dynamical friction (DF), is able to flatten the inner cusp of the density profiles. We moreover assume that haloes form primarily  through quiescent accretion, in agreement with the spherical collapse  model (SCM)-secondary infall model (SIM) prescriptions. The results of this paper follow from the two assumptions above. 
Concerning the MSP and TBTF problem, applying to the Via Lactea II  (VL2) subhaloes a series of corrections similar to those of Brooks et al. (2013),  namely applying a Zolotov et al. (2012)-like correction obtained with  our model, and further correcting for the UV heating and tidal  stripping, we obtain that the number of massive, luminous satellites  is in agreement with the number observed in the MW. The model also  produces an angular momentum distribution in agreement with  observations, that is with the distribution of the angular spin  parameter and angular momentum of the dwarfs studied by van den Bosch,  Burkert, \& Swaters (2001). In conclusion, the small scale problems of  the CDM model can all be solved by introducing baryon physics.
}

\begin{document}
\maketitle

\section{Introduction}

Despite the $\Lambda$CDM model being highly successful in describing the observations of the Universe, its large scale structure and evolution (Spergel et al. 2003, Komatsu et al. 2011; Del Popolo 2007, 2013, 2014), it has some problems in describing structures at small scales (e.g., Moore 1994; Moore et al. 1999; Ostriker \& Steinhardt 2003; Boylan-Kolchin, Bullock, and Kaplinghat 2011, 2012;  Oh et al. 2011)\footnote{Other problems of the $\Lambda$CDM model are the cosmological constant problem (Weinberg 1989; Astashenok, \&Del Popolo 2012), and the ``cosmic coincidence problem".}. The main problems are the a) cusp/core problem (Moore 1994; Flores \& Primak 1994), namely the discrepancy between the flat density profiles of dwarf galaxies and LSBs and the cuspy profile predicted by dissipationless N-body simulations (Navarro, Frenk \& White 1996, 1997; Navarro 2010); b) the ``missing satellite problem" (MSP), namely the discrepancy between the number of predicted subhalos in N-body simulations (Moore et al. 1999) and the one observed\footnote{For the Milky Way (MW) the difference is larger than an order of magnitude.}; c) the angular momentum catastrophe (van den Bosch, Burkert,\& Swaters, 2001 (VBS)) namely the angular momentum loss in Smooth Particle Hydrodynamics (SPH) simulations of galaxy formation, giving rise to disks (of dwarf galaxies) having angular momentum distributions different than those of cold dark matter haloes, and disc size of simulated galaxies much smaller than the real ones.

Concerning the Cusp/Core problem, numerical simulations predict an inner cuspy profile, with inner slope $\rho \propto r^\alpha$, with $\alpha=-1$, from dwarf galaxies to clusters of galaxies, in the case of the Navarro, Frenk, \& White (1996, 1997) (hereafter NFW) profile, $\alpha=-1.5$ in the Moore et al (1998), and Fukushige \& Makino (2001) simulations, or different values according to the objects considered 
%while a series of other papers queryed the universality of the NFW density profile 
(Jing \& Suto 2000; Ricotti 2003; Ricotti \& Wilkinson, 2004; Ricotti  Pontzen, \& Viel 2007; Del Popolo 2010 (DP10), Del Popolo 2011 (DP11); Cardone \& Del Popolo 2012; Del Popolo, Cardone, \& Belvedere 2013). 
More recently (Stadel et al. 2009; Navarro et al. 2010), it was shown that a better fit to density profiles is the Einasto profile, characterized by a profile that flattens (becomes shallower) going to the center, until values of $\alpha \simeq 0.8$ at 120 pc (Stadel et al. 2009). The previously quoted smallest value for the slope obtained in simulations is larger than that in observations, which usually show flat or almost flat galaxy inner density profiles (Burkert 1995; de Blok, Bosma, \& McGauch 2003; Swaters et al. 2003; Del Popolo 2009 (DP09) , Del Popolo 2012a,b (DP12a, DP12b); Del Popolo \& Hiotelis 2014; Oh et al. 2011; Oh et al. 2010, 2011; Kuzio de Naray \& Kaufmann 2011). The cusp/core problem, initially noticed in galaxies, is a problem also present in clusters of galaxies. The studies by Sand et al. (2004), and more recently those of Newman et al. (2009, 2011, 2013), found flat inner density profiles in the case of several clusters. Several solutions to the problem have been proposed (see de Blok 2010 for a review). 
%
%%Historically, it was proposed that the problem was due to observations which would have had poor resolution or to the fact that systematic effects, %%(off-centering, non-circular motions, beam smearing) were not properly taken into account (van den Bosch, Robertson, and Dalcanton 2000; van den Bosch, %%\& Swaters  2001 (VS01)). A way to reconcile simulations results and observation proposed by Hayashi et al. (2004) purported the idea that the %%discrepancy was connected to the fact that was not correct to compare rotation speeds of gaseous disks to circular velocities obtained from a spherical %%average of the 3D DM haloes. Another idea was that simulations had not enough resolution, and were affected from overmerging and over-relaxation?? %%problems (Taylor, Silk, \& Babul 2004). Nowdays, it is well known that high resolution observations are able to distinguish between cusps and cores %%(Kuzio de Naray and Kaufmann 2011 aggiundgere), and that dissipationless have enough resolution and are not affected by overmerging or other systematic %%problems\footnote{Diemand, Moore, \& Stadel, (2004), showed through convergence tests that CDM density profiles are correctly calculated by %%dissipationless simulations. }.
%%However, a big problem remains, dissipationless simulations are not taking into account baryons, which are not negligible in the inner kpc of galaxies %%and in the inner 10 kpc of clusters, where they dominate over DM (Sand et al. 2004; Newman et al. 2009, 2011, 2012). 
%

The solutions to the cusp/core problem can be distinguished (similarly to the solutions to other small scale problems) into cosmological and astrophysical solutions. 
Cosmological solutions are based on modifying the power spectrum at small scales (e.g. Zentner \& Bullock 2003), or modifying the constituent particles of DM (Colin, Avila-Reese \& Valenzuela 2000; Sommer-Larsen \& Dolgov 2001; Hu, Barkana \& Gruzinov 2000; Goodman 2000; Peebles 2000; Kaplinghat, Knox, \& Turner, M. S. 2000). Otherwise, modified gravity theories, like $f(R)$ (Buchdal 1970; Starobinsky 1980), $f(T)$ (see Ferraro 2012), and MOND (Milgrom 1983a,b), can solve the problem.  

The astrophysical solutions are based on the idea that some ``heating" mechanism produces an expansion of the DM component of the galaxy, reducing the inner density. In this group, the main categories are: ``supernova-driven flattening" (Navarro et al. 1996; Gelato \& Sommer-Larsen 1999; Read \& Gilmore 2005; Mashchenko et al. 2006, 2008; Governato et al. 2010; Pontzen \& Governato 2011), and dynamical friction from baryonic clumps 
%or disk instablities 
(El-Zant et al. 2001, 2004; Romano-Diaz et al. 2008, 2009; Del Popolo 2009; Cole et al. 2011).

Concerning the second problem of the $\Lambda$CDM model, the MSP, Klypin et al. (1999), and Moore et al. (1999) noticed that numerical simulations predicted much more subhaloes in galactic and cluster haloes. The number of satellites having circular velocities larger than Ursa-Minor and Draco were $\simeq 500$, while as it is well known the MW dSphs are much less (9 bright dSphs (Boylan-Kolchin, Bullock, and Kaplinghat 2012), Sagittarius, the LMC and the SMC). Subsequent cosmological simulations confirmed the problem (Aquarius, VL2, and GHALO simulations (Springel et al. 2008; Stadel et al. 2009; Diemand et al. 2007b)). The discovery of the ultra-faint MW satellites (Willman et al. 2005; Belokurov 2006; Zucker 2006; Sakamoto \& Hasegawa 2006; Irwin et al. 2007) mitigated the problem but it did not completely solve it.

The MSP recently has shown another feature, while analyzing the Aquarius and the Via Lactea simulations. Simulated haloes have $\simeq 10$ (Boylan-Kolchin, Bullock, and Kaplinghat 2011, 2012) subhaloes too massive and dense to host MW brightest satellites. While $\Lambda$CDM simulations predict at least 10 subhaloes having $V_{max}> 25$ km/s, the dSphs of the MW all have $12  <V_{max}< 25 $ km/s, implying a discrepancy among simulations and the dynamics of the MW brightest dSphs (Boylan-Kolchin, Bullock, and Kaplinghat 2011, 2012). This feature of the MSP problem has been dubbed the Too-Big-To-fail (TBTF) problem\footnote{"Too big to fail" means that the simulation satellites are too big with respect to MW satellites, and there is no way to have them remain invisible.
%(to hide them).
}.

Also in this case, two main classes of solutions have been proposed: cosmological and astrophysical. Similarly to the case of the cusp/core problem, cosmological solutions are based on modification of the spectrum of perturbations or of the particles constituting dark matter. Astrophysical solutions are connected to baryon physics, and consider: a) the change of shape of satellites from cuspy to cored (Zolotov et al. 2012 (Z12); Brooks et al. 2013 (B13)), which makes the satellites more subject to tidal stripping effects and even subject to being destroyed (Strigari et al. 2007; Pe\~narrubia et al. 2010). In this picture, the present-day dwarf galaxies could have been more massive in the past, and they were transformed and reduced to their present ``status" by strong tidal stripping (e.g., Kravtsov, Gnedin \& Klypin 2004). b) Suppression of star formation due to supernova feedback (SF), photoionization (Okamoto et al. 2008; B13), and reionization. In particular, reionization can prevent the acquisition of  gas by DM haloes of small mass, then ``quenching" star formation after 
$z \simeq 10$ (Bullock, Kravtsov, \& Weinberg 2000; Ricotti \& Gnedin 2005; Moore et al. 2006). This would suppress dwarfs formation or could make them invisible; c) effects of a baryonic disc (Z12; B13). Disc shocking to the satellites passing through the disk produce strong tidal effects on satellites, stronger if the satellite has a cored inner profile. 

In the case of the TBTF problem other solutions have been proposed. The excess of massive subhalos in MW could disappear if satellites density profiles are modeled through Einasto's profiles, or if the MW's virial mass is $\simeq 8 \times 10^{11} M_{\odot}$ instead of $\simeq 10^{12} M_{\odot}$ (Vera-Ciro et al. 2012; Di Cintio et al. 2013).

The last problem is the ``angular momentum catastrophe". 
%
%%The standard model of disk formation is based on a series of assumptions: 
%%a) angular momentum originates from tidal torques (Hoyle 1953; Peebles 1969; White 1984; Ryden 1988; Eisenstein \& Loeb 1995; %%Catelan \& Theuns 1996), and is conserved while baryons give rise to the disk (Mestel 1963); b) initially DM and baryons have %%the same distribution (Fall \& Efstathiou 1980; van den Bosch F. C., Abel T., C. C. R. A., Hernquist L., White S. D. M., 2002); %%c) adiabatic contraction (Blumenthal et al. 1986; Gnedin et al. 2004; Gustafsson et al. 2006);  d) realistic halo profile (Mo et %%al. 1998); e) bulge forms from disk instabilities 
%%(van den Bosch F. C. 1998); f)  
%%supernova feedback (Navarro \& Steinmetz, 2000). 
%

SPH simulations of galaxies show that the angular momentum of baryons is not conserved in the collapse, with the result that baryons have just 10\% of the angular momentum typical of real galaxies and that disks are too small in comparison with real disks (Navarro \& Benz 1991; Navarro \& Steinmetz 1997; Sommer-Larsen, Gelato,  Vedel, 1999; Navarro \& Steinmetz 2000). 
%and disk formation find that a considerably part of dark matter is trasferred to DM by baryons with the result that disks are too small in comparison %with real disks (Navarro \& Benz 1991; Navarro \& Steinmetz 1997; Sommer-Larsen, Gelato,  Vedel, 1999; Navarro \& Steinmetz 2000)
This problem is usually indicated with the name of ``angular momentum catastrophe".  
Moreover, also the specific angular momentum distribution obtained in N-body simulations disagrees with observations (mismatch of the specific angular momentum profile). 
%{\bf Other problems add to the previous: the spread in disk sizes seems to be
%narrower then the spread in the spin parameter, l , values[268]. Major mergers should lead to spheroids, but they also
%have the highest l values [269].
%}

The angular momentum catastrophe has been often connected to the over-cooling problem in CDM models (e.g., White \& Reese, 1978; White \& Frenk 1991). In absence of feedback effects heating the gas (e.g., UV background reionization, ram pressure, tidal heating), the gas contracts forming small haloes and then collapses towards the center of the system loosing all angular momentum which is transferred by dynamical friction to the DM halo (Navarro \& Steinmetz 2000). Taking into account some form of feedback, like energy feedback from supernova, reduces the problem (e.g., Sommer-Larsen, Gelato,  Vedel, 1999).
However, even if feedback can avoid the loss of angular momentum other problems remain, like the mismatch of the angular momentum profiles (e.g., VBS), or the fact that the scatter of the logarithm of the spin parameter of real galaxies is smaller than that in simulations (de Jong \& Lacey 2000). 

Maller \& Dekel (2002) proposed a model in which the DM distribution is different from that of gas, due to gas processes.
In this model, SF removes gas from small incoming haloes (giving rise to the low angular momentum component of the system) eliminating baryons with low specific angular momentum (see also Sommer-Larsen et al. 2003; Abadi et al. 2004). 

However, feedback models even in the absence of substructures, and then DF, have problems in forming bulgeless galaxies (van den Bosch et al, 2002), unless there exists ``selective outflows" of low angular momentum gas (D'Onghia \& Burkert 2004). 
%
%%However, van den Bosch et al (2002), showed that DM and gas had the same angular momentum distribution only if cooling is %%ignored, and moreover observed that a large percentage of the gas (5\%-50\%) was counter-rotating. As a result, collision among %%the gas with positive and negative angular momentum gave rise to a bulge component. Due to this problem even in absence of %%substructures, and then DF, bulge would in any case form, even in presence of feedback???
%
%
%%D'Onghia \& Burkert (2004) showed that haloes which were not involved in major mergers have not enough angular momentum to give %%rise to the observed discs. In this case it is difficult to envisage a feedback process increasing the gas angular momentum, %%except ``selective outflow" of gas having low angular momentum. 
%
Recently, Governato et al. (2010) showed that outflows from supernovae explosions, removing gas with low angular momentum produces bulge-less galaxies, with a baryonic angular momentum distribution close to that of the galaxies stellar disc, and also flat galaxy density profiles.

The previous discussion shows that the small scale problems can be solved appropriately taking into account the baryon physics. 
A trial to find a unifying baryon solutions to the small scale problems is that of Z12, and B13. The quoted papers use the idea of Governato et al. (2010).
%that SF removes low angular momentum gas with the effect of flattening the galaxy density profiles, forming bulge-less galaxies, %and giving rise to a baryonic angular momentum distribution close to that of the galaxies stellar disc. 
Z12 showed that the same model can solve the TBTF. Z12 found a correction to the velocity in the central kpc of galaxies to mimic the flattening of the cusp due to SF, and the enhancement of tidal stripping produced by a baryonic disc. This correction together with destruction effects produced by the tidal field of the baryonic disc, and the identification of subhaloes that remain dark because of inefficiency in forming stars due to UV heating, were applied by B13 to the VL2 subhaloes of the VL2 simulation (Diemand et al. 2008). As a result the number of massive subhaloes in the VL2 were in agreement with the number of satellites of MW and M31. 
%MSP REALE ???

In the present paper, we study if the cusp/core problem, the angular momentum catastrophe, the MSP, and the TBTF problem can be simultaneously solved using baryon physics, through the model of Del Popolo (2009). 
%The quoted method is an improved secondary infall model 

The paper is organized as follows. In Sect. 2, we summarize the model. Sect. 3 describes the results and discussion. Sect. 4 is devoted to conclusions. 

\section{Model}

The model used in the following was already described in Del Popolo (2009), and Del Popolo (2012a, b) (DP12a, DP12b). Here, we give a summary.  

The model discussed in DP09, DP12a, b, is an improvement to the spherical infall models (SIM) already discussed by several authors  
(Gunn \& Gott 1972; Bertschinger 1985; Hoffman \& Shaham 1985; Ryden \& Gunn 1987; Ascasibar, Yepes \& G\"ottleber 2004; Williams, Babul \& Dalcanton 2004)\footnote{Changes to the spherical collapse introduced by dark energy where studied in Del Popolo, Pace, \& Lima (2013a, b); Del Popolo et al. (2013).}.

However, while previous authors studied how the basic SIM of Gunn \& Gott (1972) is changed by introducing one effect at a time, as a) just random angular momentum (e.g., Williams, Babul \& Dalcanton 2004), b) just adiabatic contraction (e.g., Blumenthal et al. 1986; Gnedin et al. 2004; Klypin,  Zhao, and Somerville 2002; Gustafsson et al. 2006), or c) just the effect of dynamical friction of DM and stellar clumps on the halo (El-Zant et al. 2001, 2004; Romano-Diaz et al. 2008), in our model the previous effects (adiabatic contraction, dynamical friction, random angular momentum), and others (ordered angular momentum, gas cooling, star formation (see the following)) are all simultaneously taken into account.
%the effect of adiabatic contraction, dynamical friction, ordered and random angular momentum, are all simultaneously taken into account......... 

{As already reported, Gunn \& Gott (1972) studied the self-similar collapse in an expanding universe, until ``shell crossing"\footnote{After a shell reaches the turn-around radius it starts to collapse encountering the inner shells which are expanding or recollapsing giving rise to crossing of distinct shells.}. The evolution after ``shell crossing" was studied by Gunn (1977).

In the SIM, a spherical perturbation is divided into spherical shells and its evolution followed in time. 
Since the initial density is larger than the critical one, each particle's trajectory oscillate through the center of symmetry.  
If $x_i$ is the initial comoving radius, each shell, and the particles in it, will expand to a maximum radius, $x(t, M_{\rm i})$, dubbed apapsis, different for each shell. The first apapsis passage is named turn-around, radius, $x_{ta}$, or $x_m$.
%from its initial comoving radius $x_i$ to the turn-around radius, $x_{ta}$, or $x_m$. 
After turn-around and collapse, the trajectory oscillates through the symmetry center, shell crossing starts and the mass interior to a shell is no longer constant, together with the energy. In order to handle the effect of shell-crossing and make the problem analytically tractable an adiabatic invariant is introduced (Gunn 1977). Gunn (1977) arrived to the conclusion of such an adiabatic invariant observing that the ratio of apapsis distance to the current turn-around distance decreases with time, and so does the ratio of the period of oscillation to the time scale for halo evolution (Fillmore \& Goldreich 1984). In other words, the central shells make many oscillations before there are significant changes of the potential (Gunn 1977; Fillmore \& Goldreich 1984), and the action variables\footnote{The radial action variables is given by $J_{\rm r}= 2 \pi \int_{r_{\rm a}}^{r_{\rm b}}
(dr/dt)dr$, and the angular action is the angular momentum.} associated with the particle's motion, in a potential which slowly varies, are invariant. Concerning the ``goodness" of the assumption indirect proofs are the good agreement between the results of 
high-resolution numerical simulations and those of the SIM (e.g., Ascasibar et al. 2007). 
}

One initially calculates the density profile at turn-around, $\rho_{ta}(x_m)$, (Peebles 1980; White \& Zaritsky 1992; Hoffman \& Shaham 1985), assuming mass conservation and no ``shell-crossing". Then after shells start to cross, the final density profile is obtained assuming that the central potential varies adiabatically (Gunn 1977, Fillmore \& Goldreich 1984), leading to
\begin{equation}
\rho(x)=\frac{\rho_{ta}(x_m)}{f(x_i)^3} \left[1+\frac{d \ln f(x_i)}{d \ln g(x_i)} \right]^{-1},
\label{eq:dturnnn}
\end{equation}
where $f(x_i)=x/x_m$ is the collapse factor,  
%(see Eq. A18, DP09). 
and the turn-around radius of a shell, $x_m$, is a monotonic increasing function of $x_i$, given by
\begin{equation}
x_m=g(x_i)=x_i \frac{1+ \overline{\delta}_i}
{\overline{\delta}_i-(\Omega_i^{-1}-1)},
\end{equation}
being $\overline{\delta}_i$ the
%{\bf the mean fractional 
density excess inside the shell (see Appendix A of DP09).

In our model, we consider systems with DM and baryons. The way in which they were introduced, and how we fixed their distribution  
in our study was discussed in Appendix E of DP09. 

{In the case of the dwarfs of VBS, the baryonic fraction was calculated by means of their data. In the case of the other objects for which we do not have direct measures 
%In studies not connected to peculiar objects (like the dwarfs of VBS), 
the baryon fraction was fixed as in 
%%%%%%%%%%%%%%%%%%%%%%%%%%%%%%%%%%%%%Giodini et al. (2009), and 
McGaugh et al. (2010). We want to stress that we did not posit two different class of objects, and the baryonic fraction obtained by VBS (see their table 2), are in agreement with those obtained by McGaugh et al. (2010) (see their table 2).}

The detected baryonic fraction, $f_d$, is given by 
\begin{equation}
f_d = (M_b/M_{500})/f_b=F_b/f_b
\end{equation}
where $F_b=M_b/M_{500}$ is the baryonic fraction, and $f_b=0.17 \pm 0.01$ (Komatsu et al. 2009) is the universal baryon fraction.
%As shown from McGaugh et al. (2010) (see their Fig. 2), the fraction of baryons detected in all forms deviates monotonically from %the cosmic baryon fraction as a function of mass. On the largest scales of clusters, most of the expected baryons are detected, %while in the smallest dwarf galaxies, fewer than 1\% are detected. 
%where $f_b=0.17 \pm 0.01$ (Komatsu et al. 2009).

The virial masses, 
%$M_{200}$, and 
$M_{vir}$, are converted to $M_{500}$ using the method used in White (2001), Hu \& Kravtsov (2003), and Lukic et al. (2009). 
In the case of the VBS dwarfs, the baryonic fraction was calculated by summing the disc mass, $M_{disc}$ and the gas mass, $M_{gas}$, by them calculated and by the previous mass for the virial mass, $M_{vir}$, estimated from their data (and converted to $M_{500}$)\footnote{Usually structures are labeled by the their density contrast with respect to the critical density, $\rho_c$. The mass in a given radius encompassing the overdensity $\Delta$ is given by
\begin{equation}
M_\Delta=\frac{4 \pi}{3} \Delta \rho_c R_\Delta^3,
\end{equation}
$M_{500}$ is the mass enclosed in $R_{500}$, defining the radius within which the mean structure overdensity is 500 times the critical density $\rho_c$.}.

During the evolution of the perturbation and its collapse, a random angular momentum $j(r,\nu)_{rand}$, is generated by random velocities (Ryden \& Gunn 1987). 
Usually, in the SIM papers in which angular momentum is taken into account, only the random angular momentum is considered and assigned at turn-around
(Nusser 2001, Hiotelis 2002; Le Delliou \& Henriksen, 2003; Ascasibar, Yepes \& G\"ottleber 2004), as
\begin{equation}
j_{rand}=j_{\ast} \propto \sqrt{GM r_m},  
\end{equation}
or obtained by the random inner motions to the proto-structure (Ryden \& Gunn 1987; Avila-Reese et al. 1998; Williams et al. 2004; Del Popolo \& Kroupa 2009).    
Instead of directly assigning $j_{rand}$, one can express it in terms of the ratio of pericentric, $r_{min}$, to apocentric radii, $r_{max}$, $e=\left( \frac{r_{min}}{r_{max}} \right)$ (Avila-Reese et al. 1998), 
which is constant, and approximately 0.2 in N-body simulations\footnote{A value $e \simeq 0.2$ produces density profiles close to the NFW model. In Avila-Reese et al. (1998, 1999) it was fixed at 0.3.}.
%Avila-Reese et al. (1998) used $e_0$ as a free parameter to take into consideration processes related to mergers and tidal forces that could produce %tangential perturbations. In their paper, they showed that the detailed description of these processes is largely erased by the virialization process, %remaining only through the value of $e_0$, which they fixed to $e_0=0.3$. The value $e \simeq 0.2$ gives density profiles very close to the NFW profile %(Avila-Reese et al. 1998, 1999). The previous procedure is based on results of N-body simulations of CDM halo collapse, giving constant 
%$< \frac{r_{min}}{r_{max}}> \simeq 0.2$ ratios of dark matter particles in virialized haloes.
A more detailed analysis shows that particle orbits tend to become more radial when they move to the turn-around radius, and moreover eccentricity depends from the dynamical state of the system, so that 
\begin{equation}
e(r_{max}) \simeq 0.8 (r_{max}/r_{ta})^{0.1},
\end{equation}
for $r_{max}< 0.1 r_{ta}$, (Ascasibar, Yepes \& G\"ottleber 2004). Random angular momentum is modeled through the Avila-Reese et al. (1998) method with the Ascasibar, Yepes \& G\"ottleber (2004) correction.

The other form of angular momentum is the ``ordered angular momentum", $h(r,\nu )$\footnote{The peak height $\nu$ is defined as $\nu=\delta(0)/\sigma$, where $\delta(0)$ is the central overdensity value, and $\sigma$ is the density field mass variance (Eq. B12 in DP09).}, produced by the tidal torque of the large-scale structure on the proto-structure 
%imparting to angular momentum of the proto-halo 
(Hoyle 1953; Peebles 1969; White 1984; Ryden 1988; Eisenstein \& Loeb 1995; Catelan \& Theuns 1996). 
The total specific angular momentum, $h(r,\nu )$, is obtained by integrating the tidal torques, $\tau (r)$, over time (e.g., Ryden 1988, Eq. 35).
It is common to express the total angular momentum in terms of the spin parameter, previously described. If the system is constituted of DM and baryons, the DM, and baryonic spin parameter is given by 
\begin{equation}
\lambda_{gas(DM)}=\frac{L_{gas(DM)}}{M_{gas(DM)} [2G(M_{gas}+M_{DM})r_{vir}^{1/2}]}, 
\end{equation}
$M_{gas(DM)}$ being the gas(DM) mass inside, $r_{vir}$, the virial radius, and $L_{gas(DM)}$ is the angular momentum of gas(DM).  
%The ratio of $\lambda_{gas}/\lambda_{DM}$ is fixed according to G\"ottleber \& Yepes (2007) (1.23 for haloes with $M_{vir}> 5 \times 10^{14} h^{-1} %M_{\odot}$ (see their figure 5)). 
The $\lambda$ parameter distribution is well described by a lognormal distribution (e.g. Vivitska et al. 2002).
The way dynamical friction is taken into account is described in Appendix D of DP09, and its effects on structure formation is calculated, as in DP09, by introducing the dynamical friction force in the equation of motion (Eq. A14 of DP09). 

{Since dynamical friction is an important component of the model, we summarize here the model. To start with, in hierarchical models of structure formation the cosmic environment is a collisionless medium with a hierarchy of density fluctuations. Matter is organized in lumps which are part of larger groups, in a hierarchy. The gravitational field of such a system is composed of an average field, originating from the 
%{\it 
``smoothed out" distribution of mass,
%}, 
and a stochastic field generated from the local inhomogeneities. This last component of the field can be described through a probability density $W(F_{stoch})$ (Chandrasekhar \& von Newman 1942; Kandrup 1989; Antonuccio-Delogu \& Atrio-Barandela 1992), where $F_{stoch}$ is the stochastic force. 
A body of mass $M$ (e.g., galaxy) moving through a cloud of smaller lighter bodies, of mass $m$ (substructure), distributed homogeneously and isotropically is slowed down due to the so called ``dynamical friction" force. 
%
%%{\it Another equivalent way of thinking about this process is that the light bodies are attracted by gravity toward the larger %%body moving through the cloud, and therefore the density at that location increases and is referred to as a gravitational wake. %%In the meantime, the object under consideration has moved forward. Therefore, the gravitational attraction of the wake pulls it %%backward and slows it down.} 
%
A formula often used to describe this force is the Chandrasekhar's formula (Chandrasekhar 1943), and a more general treatment is that of Kandrup(1980). In the last case the force per unit mass is given by
\begin{equation}
{\bf F}= -\mu {\bf v}=- \frac{\int W(F) F^2 T(F) d^3 F}{2<v^2>} {\bf v}
\label{eq:fric}
\end{equation}
being $\mu$ the coefficient of dynamical friction, $T(F)$ the average duration of an impulse of the stochastic force, $<v^2>$ the typical velocity of a field particle. In the case of a homogeneous system, and using the virial theorem, Eq. (\ref{eq:fric}) can be re-written as
\begin{equation}
{F}= -\mu {v}=- \frac{4.44 [G M_{\rm a} n_{\rm ac}]^{1/2}}{a^{3/2} N} \log \{1.12 N^{2/3} \} v 
\end{equation}
being $n_{\rm ac}$ the field particles comoving number density, $N$ the number of field particles in the system, and $a$ the expansion parameter. 

In order to calculate the dynamical friction force it is necessary to know the number and distribution of field particles (substructure). As already reported, in the $\Lambda$CDM model structure forms in a hierarchical bottom-up way. Halos contain subhalos, and the last contains sub-subhalos (e.g., VL2 simulation). 
The subhalos cumulative mass function can be approximated as $M_{\rm sub}= 0.0064 (M_{\rm sub}/M_{\rm halo})^{-1}$ to masses of 
$M_{\rm sub} = 4 \times 10^6 M_{\odot}$ (Diemand, Kuhlen \& Madau 2007a; Hiotelis \& Del Popolo 2006; Hiotelis \& Del Popolo 2013)\footnote{The $\Lambda$CDM model also predicts the existence of microhalos (e.g., Anderhalden \& Diemand 2013).}. 
It is interesting to note that Ma \& Boylan-Kolchin (2004) showed that even only DM clumps (subhalos) are able to flatten a cuspy profile.

In our model, gas forms clumps, which exchange angular momentum through dynamical friction with dark matter. 
The mass of the clumps, $m_{clump}$, is  $m_{\rm clump} \simeq 0.01 M_{\rm halo}$, similarly to the clumps in Cole et al. (2011) simulations. Evidences of the existence of these clumps come from observation of galaxies at high redshift (van den Bergh et al. 1996; Elmegreen et al. 2004, 2009; Genzel et al. 2011) which found clumpy structures usually refereed as ``clump clusters" and ``chain galaxies". Inoue \& Saitoh (2011) showed through SPH simulations that these clumps are responsible for the cusp to core transformation in a similar fashion to what described in this paper.     

}

A fundamental ``recipe" in galaxy formation is adiabatic contraction (AC) of DM haloes, produced by the baryons condensation in the proto-structures centers. AC produces a steepening of the DM density profile (Blumenthal et al. 1986; Gnedin et al. 2004; Gustafson et al. 2006), when baryons cool dissipatively and collapse to the proto-structure center, giving rise to a final baryonic mass distribution $M_b( r)=M_{\ast}+M_{gas}$, where $M_{gas}$ is the gas mass and $M_{\ast}$ is that of stars. Dark matter particles initially located at radius $r_i$ will move to a new radius $r< r_i$,
characterized by (Ryden 1988; Flores et al. 1993)
\begin{equation}
r_i M_i (r_i)=r \left [ M_b( r) +M_{dm} ( r) \right], 
\label{eq:ad1}
\end{equation}
where $M_{dm}$ is the final distribution of DM halo particles, and $M_i (r_i)$ the initial total mass distribution, and $M_b$, as reported, the final baryonic mass distribution. 
%Eq. (\ref{eq:ad1}) can be iteratively solved for $M_{dm} ( r)$, if $M_i (r_i)$ and $M_b( r)$ are known.
%which can be solved iteratively for the final distribution of dissipationless halo particles 
%$M_{dm} ( r)$ given the initial total mass distribution $M_i (r_i)$ and the final baryon mass distribution $M_b( r)$.
%where  $M_i (r_i)$ is the initial total mass distribution, $M_b( r)$ is the final mass distribution of dissipational baryons and %$M_{dm}$ is the final distribution of dissipationless halo particles. 
If the halo particles orbits do not cross, we have 
\begin{equation}
M_ {dm} ( r)=(1-F_b) M_i (r_i).
\label{eq:ad2}
\end{equation}
Once $M_i (r_i)$ and $M_b (r)$ are given Eqs. (\ref{eq:ad1}), (\ref{eq:ad2}) can be iteratively solved to find the distribution of halo particles. 
%Baryons and halo particles are assumed to be well mixed initially (i.e., the ratio of their densities is $F_b$ through the %protostructure). 
Usually it is assumed that the density profile of DM and baryons is the same (Mo  et al. 1998;  Keeton 2001; Treu \& Koopmans 2002; Cardone \& Sereno 2005), and it is given by a NFW profile. The final distribution of baryons is assumed to be a disk (for spiral galaxies) (Blumenthal et al. 1986; Flores et al. 1993; Mo et al. 1998; Klypin et al. 2002; Cardone \& Sereno 2005)\footnote{In DP09, the final baryons distribution of typical spiral galaxies (e.g., MW, and M31) was determined using the Klypin et al. (2002) model (see their subsection 2.1), and in the case of elliptical galaxies and clusters it was assumed an Hernquist model  (Rix et al. 1997; Keeton 2001; Treu \& Koopmans 2002).}.

The previous model was improved following Gnedin et al. (2004), who showed that numerical simulation results are better reproduced if one assumes the conservation of the product of the mass inside the orbit-averaged radius, and the radius itself
\begin{equation}
  M(\bar{r})r= {\rm const}.
  \label{eq:modified}
\end{equation}
where the orbit-averaged radius is
\begin{equation}
  \bar{r} = {2 \over T_r} \int_{r_{min}}^{r_{max}} r \, {dr \over v_r},
\end{equation}
where $T_r$ is the radial period\footnote{Gnedin et al. (2004) used $M(\bar{r})r= {\rm const}$ instead of $M(\bar{r}) \bar{r}= {\rm const}$ because the first one is a better approximation to their simulations result.}.

The previous model does not contemplate exchange of angular momentum among baryons and DM. This is a good approximation in the early phase of the structure formation, when baryons have a density an order of magnitude smaller than that of DM, 
but later, when baryons density increase because of the collapse, the approximation is no longer valid. 
Baryons density increase acts as a coupling process of DM and baryons (Klypin et al. 2001; Klypin,  Zhao, and Somerville 2002). Excitation of spiral waves and/or the presence of bar like modes can give rise to a non-axisymmetric component.
%moreover {\bf non-axisymmetric component may develop due to the excitation of spiral waves and/or bar-like modes. }
The coupling effect is very powerful in the last period of structure formation, with a reduction of DM density by a factor 
%At later stages of collapse, when baryon density increases, the effect is potentially quite large resulting in a decrease of the %dark matter density by a factor 
of ten (Klypin, Zhao, \& Somerville 2002; DP09; DP12a, b). 

Similarly to other semi-analytical models, following the philosophy of the White \& Rees (1978) and White \& Frenk (1991) works, 
we include other important physical processes such as gas cooling, and star formation (see the following).
% VEDERE SE RIMETTERLO
%%\footnote{For what concerns supernova feedback have a lok to Sect.... }.

The structure formation in our model can be summarized as follows. The galaxy formation starts with the proto-structure in its linear phase, containing DM and gas. In our model the baryons are initially in the form of a diffuse gas, with the previously quoted baryonic fraction. In order to follow the structure formation, we divide the proto-structure into mass shells, made of DM and baryons. The proto-structure evolution is followed in the expansion phase, until turn-around, and in the following collapse by means of the SIM. As is known, DM collapse earlier than baryons to form the potential wells in which baryons fall, and are subject to radiative processes with the formation of clumps, which collapse to the halo center condensing into stars, as described in Li et al. (2010) (Sect. 2.2.2, 2.2.3), De Lucia \& Helmi (2008), respectively. 
During the baryons infall phase, DM is compressed (AC). 
%by baryons infall (AC). 
At this epoch the density profile of the proto-structure steepens. In their travel towards the center, baryons clumps are subject to DF from the less massive DM particles.
This produces a predominant motion of DM particles outwards. The effect of the previous mechanism is amplified by the angular momentum acquired by the proto-structure through tidal torques (ordered angular momentum), and by random angular momentum. 
The cuspy profile is flattened to a cored one.   

{Before going on, we would like to spend some words on the reasons the model used here is valuable (a deeper discussion on this issue can be found in DP09, and DP12a). Nowadays, structure formation is mainly studied through numerical simulations. Their strength is the ability to entrap to a large extent the complexity of non-linear processes.
This is also their point of weakness because it is complicated to un-knot and interpret the dynamical
processes they take into account in terms of the implicit physics. In this respect, semi-analytic models
are much more flexible than simulations, making it possible to separate the effects of physical processes.
Structure formation in simulations develops by the aggregation and merging of subclumps of matter,
while in models like the SIM, haloes are spherically symmetric, they do not suffer major mergers, and
accretion happens in a quiescent way. Nevertheless, the SIM describes properly structure formation. A
paper of Zaroubi et al. (1996), showed that in energy space the collapse happens in a gentle way, very 
different from the chaotic collapse seen in N-body simulations. Several following papers converged on this
point (Toth \& Ostriker 1992; Moore et al. 1999; Huss et al. 1999a, b). 
Moreover, the commonness of spiral galaxies, with their thin, and fragile discs testify against the primary role of major mergers in their formation (e.g., Toth \& Ostriker 1992). This point of view is confirmed by recent simulations that showed that at high redshift disc galaxies form through smooth accretion (Dekel et al. 2009). 
Moreover, several studies compared the SIMs results against
high-resolution numerical simulations (e.g., Ascasibar et al. 2004, 2007), and showed that density profiles obtained by the SIM are in good agreement with N-body simulations.
Furthermore, particle discreteness effects in dissipationless N-body simulations has been long 
debated (see Romeo et al. 2008), and even if nowadays these effects are ``under control" (see Diemand
et al. 2004) the quoted simulations do not incorporate baryons. Also SPH simulations are not devoid
from problems. A recent paper of Marinacci et al.(2013) shows that the high resolution SPH 
simulations of the formation of spiral galaxies have several drawbacks, starting from a not clear understanding
of the details of the physics, to modeling and numerics involved in those calculations. 
%For example, slight difference in the details of the feedback implementation can give rise to totally different results,
%or changes in the resolution can give rise to galaxies with morphology varying in the entire Hubble type
%range (Okamoto et al. 2005). 
Consequently alternative approaches (e.g., the SIM) may assume an important role. 
%In the peculiar case of our model, it has shown to predict correctly, among others, the density profiles of galaxies and clusters %(DP09; DP12a,b), correlations observed (DP12a), and the mass dependence of the inner slope of halos profiles (DP10, DP11) (see the
%last part of section 5).

Our model, is able to deal with the baryonic processes shaping the inner structure of clusters (and
galaxies). It was able to predict in advance of SPH simulations (e.g., Governato et al. 2010, 2012; Martizzi et al. 2012) the correct shape of the density profiles of clusters (DP12a) and galaxies (DP09; Del Popolo \& Kroupa 2009; DP12b), and correlations among several quantities in clusters of galaxies (DP12a) later observed in Newman et al. (2013a,b). The model also predicts correctly that the inner slope of the density profiles depends on the halo mass (DP10, DP11), later seen in the SPH simulations of Di Cintio et al. (2013, 2014).

}

%TOLTO
%As shown in DP09, DP12a, b, the density profile that the previous model produces are in agreement with the El-Zant et al. (2001, %2004), Romano-Diaz et al. (2008), and Cole et al. (2011) studies, based on the role of baryonic clumps and dynamical friction on %cusp flattening in galaxies and clusters of galaxies. The density profiles is also in agreement with the Governato et al. (2010) %(see DP11 Fig. 4), and Mashchenko et al. (2005, 2006) simulations, in which the flattening of the cusp is produced by SF. 

\section{Results and discussion}

As discussed in the Introduction, the small scale problems of the $\Lambda$CDM model can be solved using cosmological or astrophysical solutions. 
The astrophysical solutions based on the role of baryons in structure formation, are more easy to constrain than cosmological solutions, and moreover do not request one to reject the $\Lambda$CDM paradigm for a new one. 

In this section, we show how the model previously discussed changes the density profile, the angular momentum distribution in galaxies, and the 
galactic subhaloes distribution and characteristics. As done in previous papers, DP09, DP12a, DP12b, we will study the evolution of protostructures from the linear phase until they form structures with galactic mass, and then we calculate their density profiles, and the DM and baryon angular momentum distribution. Then we find, similarly to Z12, an analytical correction to apply to the center of the haloes which mimics the effect of flattening of the cusp, and finally similarly to B13, we apply the previous correction together with the tidal destruction and UV heating effects on subhaloes, to the VL2, as done by B13.    

\subsection{Cusp/core problem}

\begin{figure}
\hspace{-1.5cm}
\psfig{file=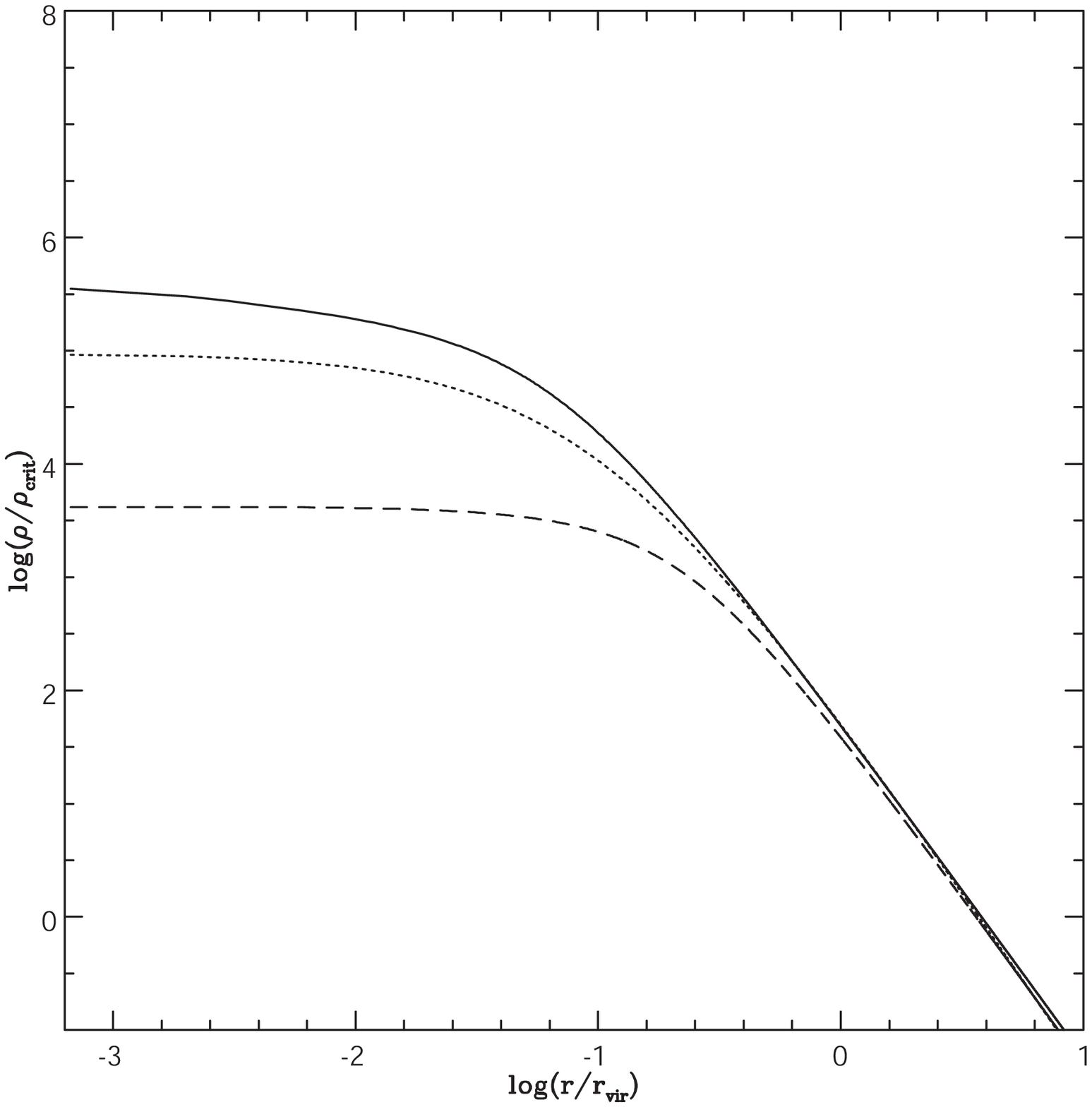,width=10.0cm}
\psfig{file=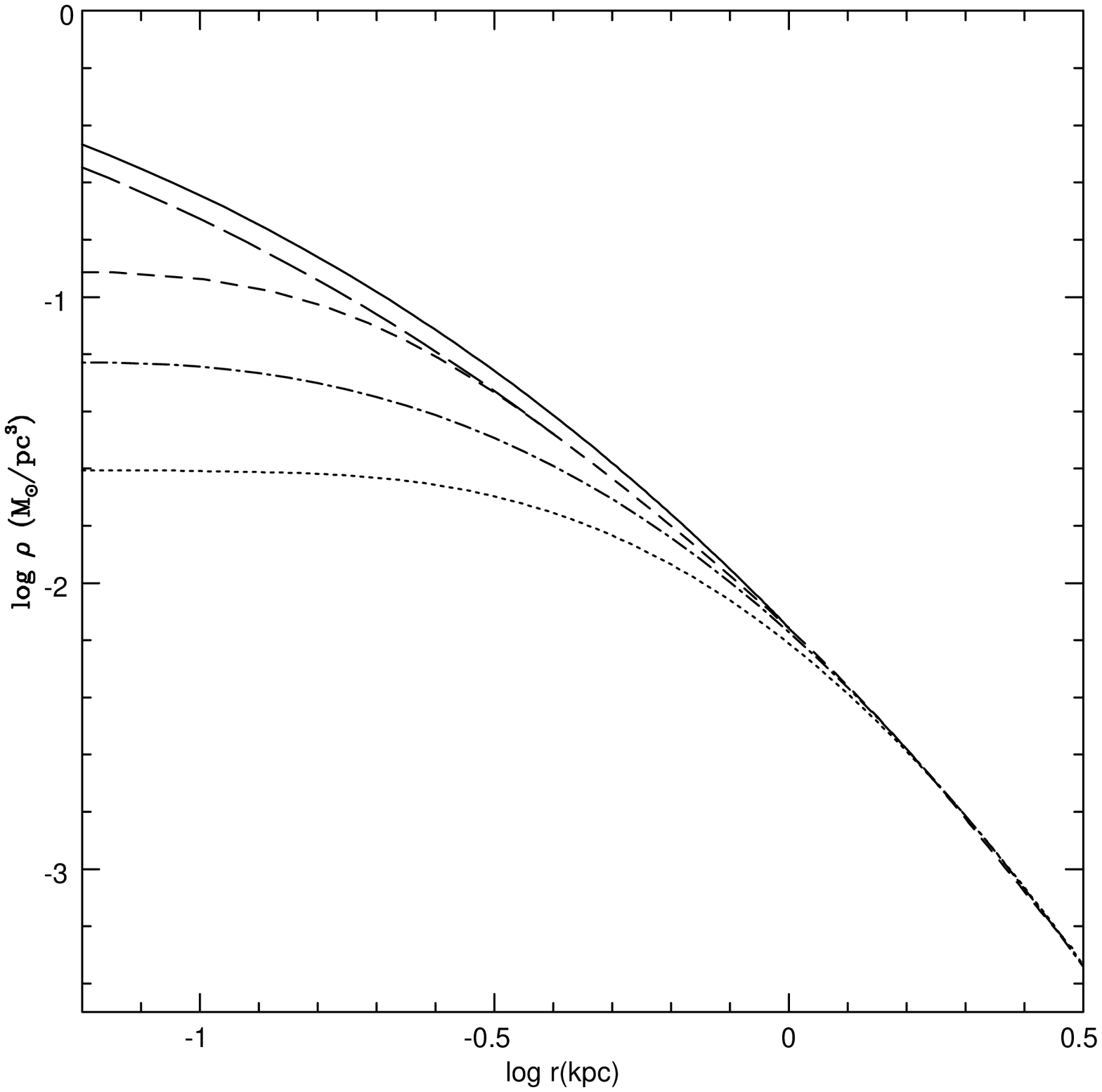,width=8.8cm}
%\picplace {2.0cm}
\caption[]{Fig 1a: density profiles of galaxies with mass $10^{8} M_{\odot}$ (dashed line), $10^{9} M_{\odot}$ (dotted line), and 
$10^{10} M_{\odot}$ (solid line). The halo with mass $10^{9} M_{\odot}$  has a baryon fraction $f_d \simeq 0.04$ and a specific angular momentum $h \simeq 400$ kpc km/s ($\lambda \simeq 0.05$), as in the case of UCG 6446.
%? (le altre?).   
Fig. 1b: density profile evolution of a $10^9 M_{\odot}$ halo starting from $z=10$ (solid line), when the profile is a NFW profile. 
The following evolution represented by the long-dashed line, short-dashed line, dot-dashed line, dotted line, represents the profile at 
$z=3, 2, 1, 0$, respectively.}
\end{figure}

\begin{figure}
\psfig{file=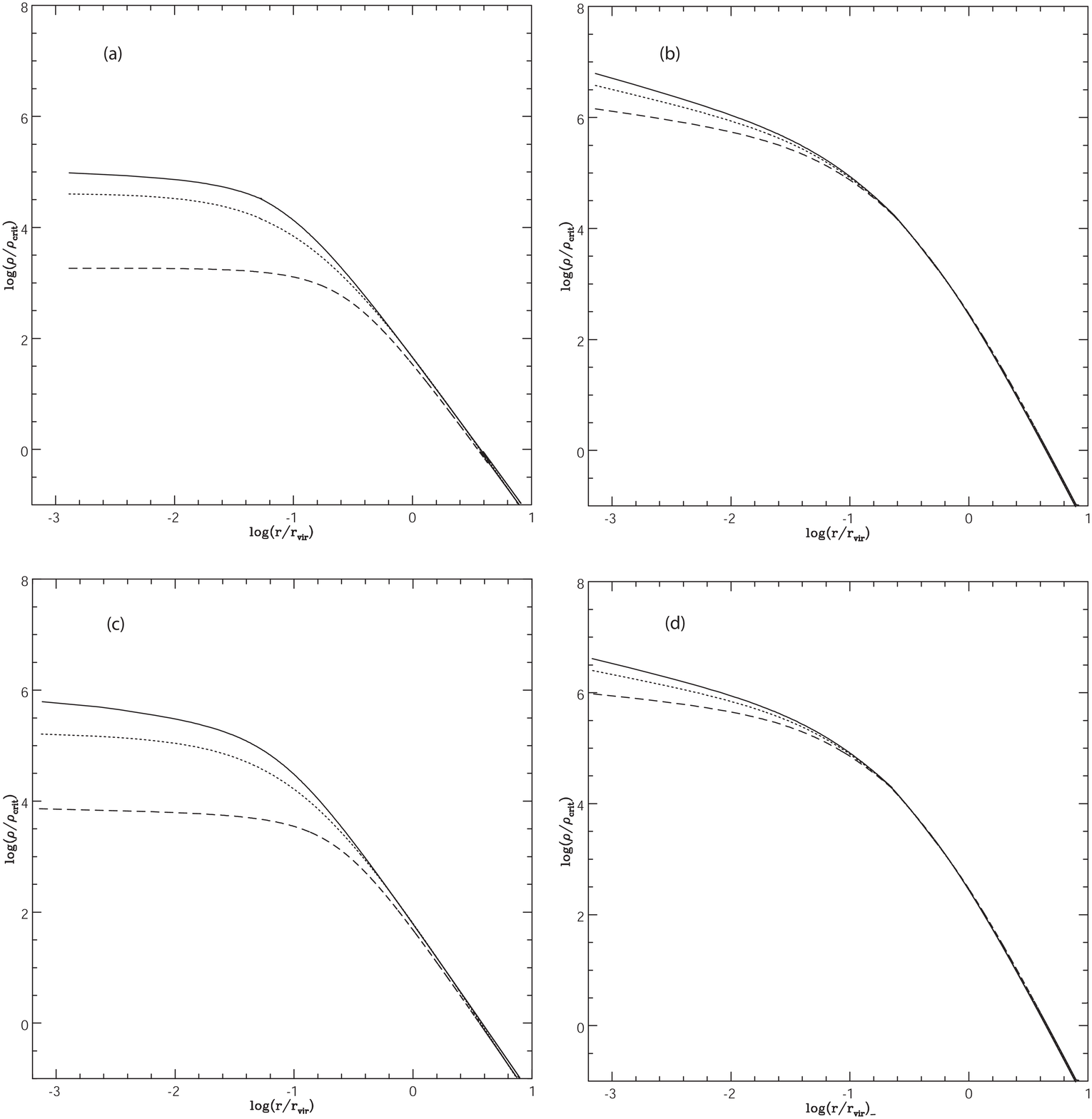,width=15.0cm} 
%\hspace{-1.5cm} 
%\psfig{file=dwarfsnbbb.eps,width=10.0cm} 
%\psfig{file=dwarfsnnn.eps,width=10.0cm} 
%\hspace{-3.5cm}
%\psfig{file=dwarfsn_fric.eps,width=10.0cm} 
%\psfig{file=dwarfsnnn_fric.eps,width=10.0cm}
\caption{Dependence of the density profiles shape on angular momentum and dynamical friction. The symbols are like in Fig. 1.
Starting from Fig. 1a, the value of the specific ordered angular momentum is changed from $h$, to $2 h$ (panel a), and from $h$, to $h/2$ (panel b), leaving $\mu$ unchanged. The value of $\mu$ is changed to $2 \mu$ (panel c), and to $\mu/2$ (panel d), leaving $h$ unchanged.
}
\end{figure}

We start with the Cusp/core problem. 
In Fig. 1a, we plot the density profile of a galaxy with $M = 10^{8} M_{\odot}$ (dashed line), $10^{9} M_{\odot}$ (dotted line), 
and $M=10^{10} M_{\odot}$. The specific angular momentum was chosen similar to one of the VBS galaxies, namely UGC 6446, to be
%57399 (6646), 
having $h \simeq 400$ kpc km /s ($\lambda \simeq 0.05$), and the baryon fraction (see Sec. 2) is $f_{d} \simeq 0.04$. The final density profile is well described by a Burkert profile
\begin{equation}
\rho(r)=\frac{\rho_0}{(1+r/r_0) [(1+(r/r_0)^2]}
\end{equation}
being $r_0$ and $\rho_0$, the scale radius, and the central density, respectively. 

A comparison of the density profiles produced by the model with the Governato et al. (2012) result, showing a good agreement with SPH simulations, was carried out in DP12a.

As Fig. 1b shows, the density profile shape is dependent on mass. The slight steepening of the density profile with mass was explained in DP09, DP10, DP12a. In summary, objects less massive originate from a smaller peak height $\nu$.
Since, the specific ordered (h), and random ($j_{rand}$), angular momentum acquired by the protostructure are anticorrelated with 
$\nu$, smaller objects acquire larger angular momentum. ``Particles" of the protostructure having larger angular momentum, will remain for a longer time close to maximum radius, giving rise to shallower profiles (see also Williams et al. 2004). As described in DP12a,b, there exists a correlation between the angular momentum acquired by the structure, the baryon content of the structure, and the shape of the density profile. 
{Then the larger is the angular momentum acquired by the galaxy, the flatter is the density profile (see also DP09; DP12a). }

The steepening of the density profile with mass is in agreement with density profiles of THINGS galaxies (de Blok et al. 2008). de Blok et al. (2008) found that brighter, and larger galaxies, with $M_B<-19$ have density profiles well described both from cored and cuspy profiles, while cored profiles are preferred for less bright, and less massive galaxies, with $M_B>-19$.

Fig. 1b, shows the evolution of a $10^9 {M_{\odot}}$ halo starting from the redshift of virialization $z=10$ (solid line). The density profile at $z=3$, 2, 1, and 0, is represented by the long-dashed line, short-dashed line, dot-dashed line, and dotted-line, respectively.  
The proto-structure and profile evolution was summarized in the final part of Sect. 2. After virialization the density profile evolution is due to secondary infall, two-body relaxation, interplay between DM and baryons. 
%{\bf The evolution after virialization is produced by secondary infall, two-body relaxation, dynamical friction, and angular %momentum. The cusp is slowly eliminated and within 1 kpc a core forms. The previous result is similar to what found by Romano-
%Diaz et al. (2008) who studied the DM cusp evolution using N-body simulations with and without baryons. The "erasing"
%of the cusp is associated by them to the heating up of the cusp region via dynamical friction (EZ01) or influx of subhalos
%into the innermost region of the DM halo. In our model, the erasing of the cusp is connected to the joint effect of dynamical
%friction and angular momentum.}
As shown in Fig. 1b, the profile flattens and a core is formed within 1 kpc. A similar result was obtained by Romano-Diaz et al. (2008), who performed N-body simulations of the DM density profile in the absence and in the presence of baryons. Cusp erasing was connected by them to an influx of sub-haloes in the central part of the halo, as well as to heating up of DM through DF, similarly to the El-Zant et al. (2001, 2004) results. 
{Concerning the model proposed by El-Zant, it is based on baryonic clumps which exchange angular momentum through dynamical friction with DM. As previously reported, there are observational evidences on the existence of these clumps, coming from high-$z$ observations of galaxies (van den Bergh et al. 1996; Elmegreen et al. 2004, 2009; Genzel et al. 2011). These clumps are of fundamental importance in the formation of bulges, in the so called ``clump-origin bulge" (Noguchi 1998, 1999; Inoue \& Saitoh 2012). Inoue \& Saitoh (2011), showed through SPH simulations that these clumps are able to flatten the cuspy profile. 

Theoretical results in agreement with the previous ones are those of Ma \& Boylan-Kolchin (2004), Nipoti et al. (2004); Romano-Diaz et al. (2008, 2009)); Cole et al. (2011). Cole et al. (2011) showed that the mechanisms described by the previous quoted authors are very efficient in cusp erasing. 
}

~\\

{We hinted to the role of angular momentum on the density profiles shapes. Another quantity on which the model depends is 
the ``magnitude" of dynamical friction. The dynamical friction force can be written (see Eq, D5 in DP09) as
\begin{equation}
F=-\mu v
\end{equation}
where $\mu$ is the coefficient of dynamical friction. The effect of changing this quantity is similar to changing the 
magnitude of the angular momentum. Namely, increasing $\mu$ gives rise to shallower profiles.  

In Fig. 2, we plot how the density profiles change when $h$, and $\mu$ are changed. Fig. 2a plots the density profiles in Fig. 1a when $h$ is increased to $2 h$, leaving $\mu$ unchanged. In Fig. 2b $h$ is decreased to $h/2$. As already reported, the increase in the magnitude of the angular momentum flattens the profile, and its decrease produces a steepening of the profile. 
In Fig. 2c, the value of $\mu$ is increased to $2 \mu$, and in Fig. 2d it is reduced to $\mu/2$, leaving unchanged the angular momentum. As already reported, the effect of dynamical friction is similar to that of angular momentum, but its effect is slightly lower. 
 
}

\subsection{Angular momentum catastrophe}

\begin{figure}
\psfig{file=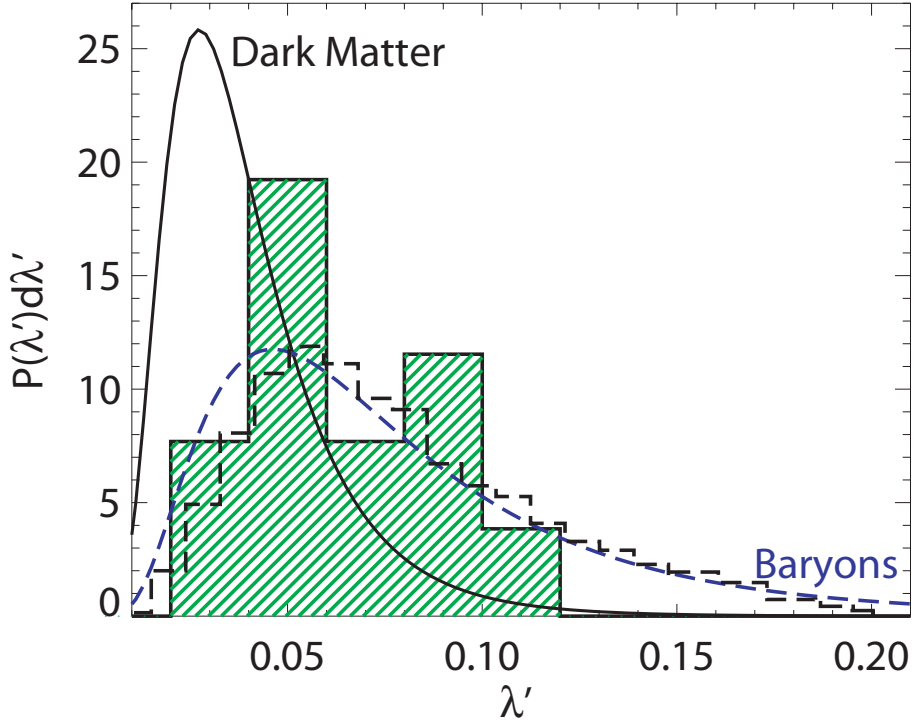,width=12.0cm}
\caption[]{
The distribution of $\lambda'$ in the sample of dwarf galaxies by VBS (shaded histogram) compared to the Maller \& Dekel (2002) predictions  
%with our model predictions with Vfb = 95 kms-1 
for the baryons in dwarf galaxies (dashed curve), and to N-body simulation result for dark halos by Bullock et al. (2000) (solid curve). 
%The inclusion of blowout produces a $\lambda'$ distribution in good agreement with the data.
The dashed histogram represents our result.
}
\end{figure}

In the following, we study the angular momentum distribution (AMD) of the structure obtained with the method described in Sect. 2 and compare it with that obtained by VBS using 14 dwarfs. 
Before, 
%discussing the VBS result, 
we recall some theoretical background.

A virialized halo made of baryonic matter and DM of virial mass $M_{vir}$ has a total specific angular momentum (SAM), 
$j_{tot}=L_{tot}/M$\footnote{Note that the SAM j used here is the total SAM, sum of h, and $j_{rand}$ seen in Sect. 2.}, $L$ being the angular momentum, proportional to the 
maximum SAM, $j_{\rm max}$, and the normalized cumulative distribution, $m(j)$
\begin{equation}
j_{\rm tot} = j_{\rm max} \left[ 1 - \int_{0}^{1} m(k) {\rm d}k
\right],
% \equiv \zeta j_{\rm max}. 
\label{jtot}
\end{equation}
where $k =j/j_{\rm max}$, and 
\begin{equation}
m(j) = \int_{0}^{j} p(j) {\rm d}j.
\label{mj}
\end{equation}
$p(j) d j$ being the mass fraction having SAM in the $j$-$j+dj$ range. 
A quantity often used to express the angular momentum, $L$ of a structure, is the spin parameter
\begin{equation}
\lambda= L \sqrt{|E|}/GM^{5/2},
\end{equation}
(Peebles 1969), where E indicates the system internal energy. For practicality reasons, the previous spin parameter definition
is substituted by
\begin{equation}
\lambda =  {j_{\rm tot} \over \sqrt{2} \,  r_{\rm vir} V_{\rm vir}}, 
%\, {\cal G}.
\label{lam}
\end{equation}
where $V_{\rm vir} = \sqrt{G M_{\rm vir} / r_{\rm vir}}$, and $r_{\rm vir}$ is the virial radius. 
For a NFW profile, $\lambda'=\lambda/\sqrt{f_c}$, and $f_c$ is a function of the concentration parameter $c_{vir}$ (Mo et al. 1998). 
As shown by Bullock et al. (2000), the AMD is well described by
\begin{equation}
m(k) = {\mu k \over k + \mu - 1}.
\label{mbull}
\end{equation}
Bullock et al. (2000) showed that for a $\Lambda$CDM Universe with $\sigma_8 = 1.0$,  $h=0.7$, $\Omega_m = 0.3$,  $\Omega_{\Lambda}=0.7$, 
$\mu$ has a Gaussian distribution in ${\rm log}(\mu - 1)$\footnote{The mean is $-0.6$ and the standard deviation $0.4$. For 90\% of the haloes the value of $\mu$ is $1.06 < \mu  < 2.0$ with $\langle \mu \rangle = 1.25$.}.
The probability distribution corresponding to Eq. (\ref{mbull}) is given by
\begin{equation}
p(t) = {\xi \mu (\mu - 1) \over (\xi t + \mu - 1)^2},
\label{prob}
\end{equation}
where $t= j/j_{\rm tot}$, and $\xi$ is the ratio of the total and maximum SAM, $j_{\rm tot} \over j_{\rm max}$, 
\begin{equation}
\xi = {j_{\rm tot} \over j_{\rm max}} = 
1 - \mu \left[ 1 - (\mu - 1) {\rm ln} \left(
{\mu \over \mu - 1}\right) \right].
\label{zeta}
\end{equation}

Going back to VBS, in their study they fitted the rotation curves of 14 dwarfs (previously studied by VBS),
assuming that the galaxies are constituted by a NFW dark halo, a thin gas disc, and a thick stellar disc. They obtained the concentration parameter, $c_{vir}$, the virial velocity, $V_{vir}$, $f_{disc}=M_{disc}/M_{vir}$, and
$f_{gas}=M_{gas}/M_{vir}$. The SAM distribution 
was obtained through $m(j)=M_{disc}(r)/M_{disc}(r_{max})$, where $j=r V_c(r)$. By means of Eq. (\ref{jtot}) they got $j_{tot}$ for each galaxy and from this the disc spin parameter
\begin{equation}
\lambda_{disc}=  {j_{\rm tot} \over \sqrt{2} \,  r_{\rm vir} V_{\rm vir}}, 
%\, {\cal G}.
\label{lam}
\end{equation}
Finally, they compared the histograms of $\lambda_{disc}$ distribution for the quoted galaxies with the $\lambda$ distribution of DM haloes (that they found agreement, in contrast with Maller \& Dekel 2002, Governato et al. 2010), and the $j$ distribution, $p(t)$, that was compared to the DM distribution.  
In their analysis, they assumed that the mass-to-light ratio was 1 in the R band\footnote{In any case, the results are not sensitive to this choice, since dwarfs are DM dominated.}.

We repeated their analysis as follows. We used the gas fraction, total SAM, obtained by them, and estimated the mass of each galaxy 
from their $V_{vir}$, assuming like them that the density profile was a NFW one\footnote{For a NFW profile, the virial mass is given by $M_v=\sqrt{\frac{3}{800 \pi \rho_c G^3}} V^3_v=\sqrt{1/100} (H_0 G)^{-1} V^3_v$.
%$M_{\Delta}= (\Delta/2)^{-1/2} (GH_0)^{-1} V^3_{\Delta}$ McGaugh (apjl 708, L14-L17, sect. 2.2.3) 
}. Then, by means of the model in Sect. 2, similarly to DP12a, we ``simulated" each galaxy. Then, similarly to VBS we obtained the spin parameter distribution, and the $j$ distribution. 
Improved results could be obtained using the VBS rotation curve data, and use a Burkert profile for the DM halo. 

Like them, we calculated the cumulative AMD, $m(j)=M_{disc}(r)/M_{disc}(r_{max})$, we got through Eq. (\ref{jtot}) $j_{tot}$, and then the spin parameter. Finally, we obtained the $p(t)$ of the AMDs.

The results are reported in Fig. 3, and Fig. 4. In Fig. 3, we plotted the $\lambda'$ distribution. The solid line represents the results of Bullock et al. (2000) for the DM spin parameter distribution, while the dashed curve, the spin parameter distribution for the baryons in the 14  VBS galaxies obtained by Maller \& Dekel (2002). The solid histogram is the spin distribution obtained from the 14 VBS galaxies (see Maller \& Dekel 2002), and the dashed histogram is the spin parameter distribution obtained from the 14 VBS galaxies, using the method of this paper.
 
The plot shows that the spin distribution of DM (solid line) is different from that of galaxies (solid histogram). The spin distribution obtained by Maller \& Dekel (2002) (dotted curve), and by us (dashed histogram), are in good agreement with the observational distribution of the spin parameter in galaxies. The Maller \& Dekel (2002) feedback model produces a good agreement with observations. As reported by them, the agreement is independent of the feedback model used.
%In our case, the   

In Fig. 4, we plot the $j$ profile for UGC 6446 obtained by VBS (Fig. 4b), and the one calculated by us (Fig. 4a). 
In the figure, the solid line is the Bullock et al. (2001) average DM j-profile ($\mu=1.25$), while the shaded area represent the 
$p(t)$ of AMD for UCG 6446.
%
%%({\bf normalized to $f_{disc}/f_{bar}$}). 
The figure shows the angular-momentum profile mismatch. The dwarf angular momentum shows a deficit of angular momentum, at the high and low end of the distribution,
with respect to the DM distribution. In models like that of Maller \& Dekel (2002) or simulations like that of Governato et al. (2010), the low-j tail in the distribution is missing because feedback (blowout) selectively removes gas from small satellites, giving rise to the material of the halo having low-j.   
%
%%{\bf The high-j tail tends to be reduced in the baryons,
%%because this tail is often the result of a small satellite that comes in with its orbital angular momentum aligned with
%%the halo spin, and now has lost its gas.}  
%
\begin{figure} \hspace{-1.8cm}
\psfig{file=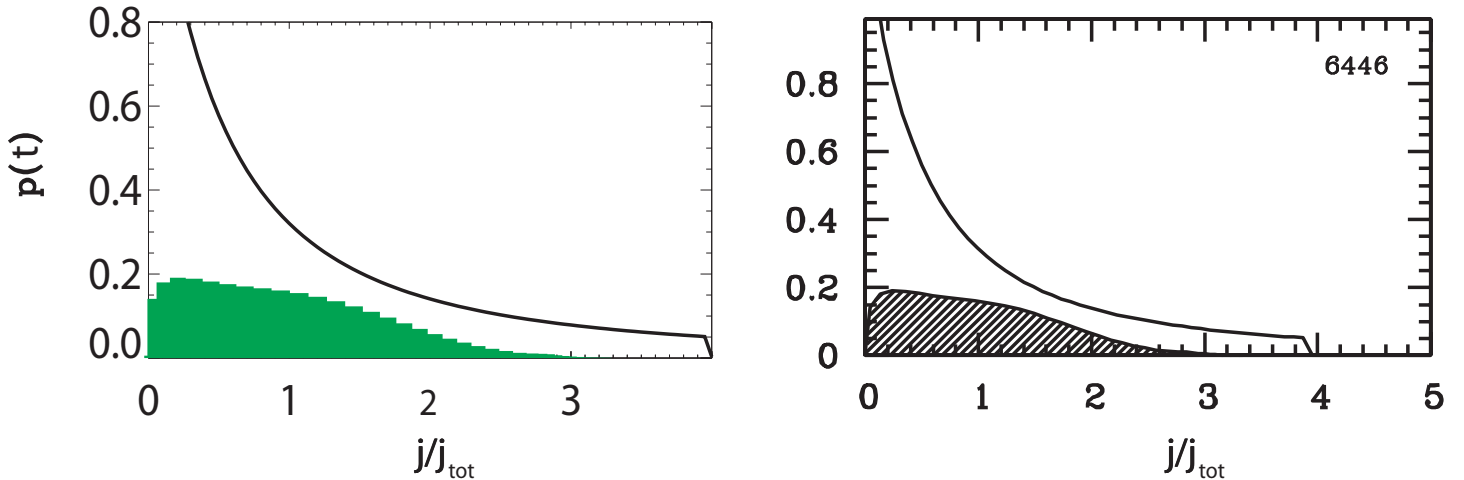,width=19.0cm}
\caption[]{The $p(t)$ of the AMDs for one of the 14 galaxies of VBS (shaded area) compared to the $p(t)$ of  
%normalized to  $f disc/ f bar$.
Eq. (\ref{prob}) with $\mu=1.25$ (normalized to unity) (solid line), namely the median of the LCDM haloes AMDs. 
In the left panel, we plot the result of our calculation, while in the right panel we plot VBS result for UGC 6446. 
The difference between the two distributions is connected to the AMD of baryons not incorporated in the disc (under the standard  hypothesis of conservation of SAM of baryons).    
Preferentially highest and lowest momenta baryonic matter is absent in the disc.
%{\bf Under the standard assumption that baryons conserve their specific angular momentum the difference between the two %distributions reflects the AMD of the baryonic matter that is not incorporated
%in the disc. Note that it is preferentially the baryonic matter with both the highest and the lowest angular momenta that is %absent in the discs. }
}
\end{figure}

\begin{figure}[ht]
\psfig{file=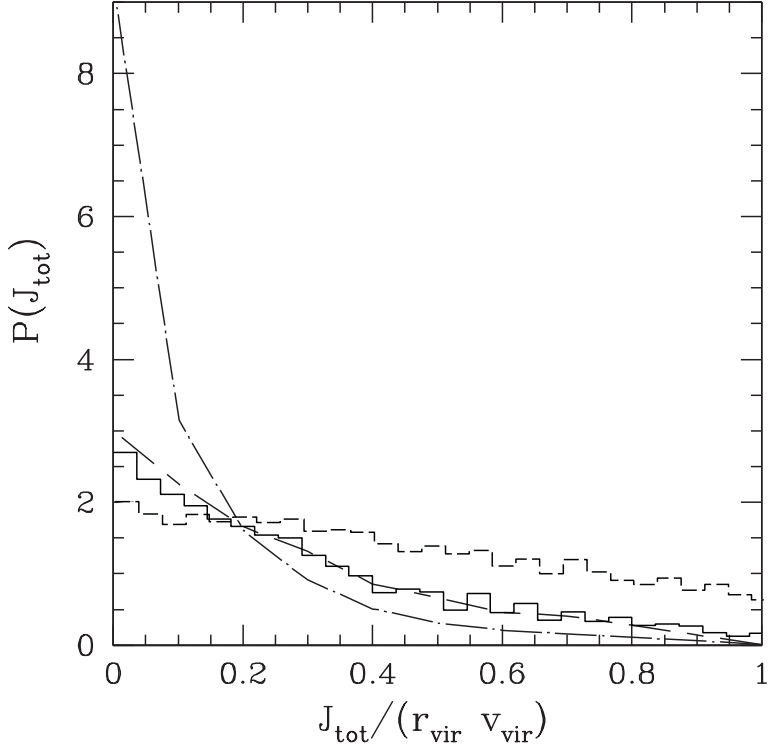,width=10.0cm}
\caption[]{Total SAM distributions. The distribution of angular momentum in the halo n. 081 (dashed line), and in the halo n. 171 (dot-dashed line) of van den Bosch et al. (2002) compared with the SAM of a $10^{12} M_{\odot}$ halo obtained with our model (dashed histogram), and that for the density profile reproducing the NFW halo (solid histogram).}
\end{figure}

{In our model, the reduction of DM density in the inner part of haloes (previously described) and the change of the angular momentum distribution, is due to the following reasons. 

%Differently from numerical simulations, 
Already more than a decade ago, semi-analytic models of galaxy and disc formation of isolated halos predicted the correct size of discs and of disc galaxy formation (e.g., Mo, Mao, \& White 1998; van den Bosch 2002;  Natarajan 1999; Firmani \& Avila-Reese 2000; Buchalter, Jimenez, \& Kamionkowski 2001; Mayer, Governato, \& Kaufmann 2008).
The models assumed that a) angular momentum is conserved during the collapse phase of baryons inside halos, that b) baryons have the same initial specific angular momentum distribution of DM. Vice versa, only recently it was understood that a necessary condition to form realistic galaxies and discs in simulations is to have high spatial and mass resolution (Mayer, Governato, \& Kaufmann 2008; Governato et al. 2010). This avoid ``spurious" dissipation of angular momentum that can be produced by the interaction of particles having noteworthy temperature difference or by artificial viscosity. Moreover, during merging angular momentum is lost through DF. As before reported, the previous problem can be solved if the thermal energy of baryons is enough to resist collapse, changing the merging hierarchy dictated by dark matter, and forming a smoother distribution. If gas is not clumpy, but has a smooth distribution, this reduces the angular momentum loss caused by DF.

In our semi-analytic model, structure formation undergoes quescient accretion, and minor mergers, differently from simulations in which structures are obtained as a result of major mergers of sub-haloes. 
The existence of many spiral galaxies with thin disks somehow supports the previous quiescent formation mechanism. Moreover, Dekel et al. (2009), {argued} that disc galaxies, at high $z$, are not formed through galactic mergers, but by smooth accretion flows. Consequently, the angular momentum loss to which simulations are exposed, is not present.  

This can be shown, as already done in DP09, and Williams et al. (2004), as follows. In Fig. 5 the dot-dashed line is the specific angular momentum distribution of the halo n. 170 of van den Bosch et al. (2002) (resembling most of the SAMs in van den Bosch et al. 2002). The dashed line represents the specific angular momentum distribution of the halo n. 81 of van den Bosch et al. (2002) (shallowest distribution found in their simulations). The dashed histogram represents the SAM of a halo of $10^{12} M_{\odot}$ obtained with our model. Starting from the $10^{12} M_{\odot}$ halo obtained in our model, we reduced the angular momentum 
%and the coefficient of dynamical friction 
in order to reproduce a NFW profile having $10^{12} M_{\odot}$, and $c=10$. Then we calculated the SAM for this halo, which is the solid line histogram in Fig. 5. This last histogram is closer to the typical halos coming out from numerical simulations than the histogram relative to our halo. Moreover, the histogram of the SAM of the halo reproducing the NFW profile (solid histogram)
is more centrally concentrated than the SAM of the halo of the initial halo (dashed histogram). Fig. 5 suggests that 
halos in simulations loose angular momentum between $0.1 r_{\rm vir}$ and $1 r_{\rm vir}$, when compared to the halos obtained in our model, or in semi-analytic models (see Williams et al. 2004), in agreement with the ``angular momentum catastrophe"  
%TOLTO
%%In fact, several authors have noted that N-body–generated dark matter halos have too little angular
%%momentum compared to the halos of real disk galaxies, possibly because it was lost during repeated collisions through
%%dynamical friction or other mechanisms (van den Bosch et al. 2002; Navarro \& Steinmetz 2000).
%

Similarly to what happened with the semi-analytic model previously cited (e.g., Mo, Mao, \& White 1998), halos and galaxies 
forming in our model are closer to real halos and galaxies than those generated in simulations, at least closer that those generated in simulations of some years ago.
%suffering of the ``angular momentum catastrophe". 
As previously reported, recently, increasing the simulations resolution and properly using feedback, simulations are able to form realist galaxies (e.g., Mayer, Governato, \& Kaufmann 2008; Governato etl a. 2010). 
% TOLTO
%%The fact that thin extended spiral disks appear to be common in the universe further supports
%%our hypothesis, since it implies that quiescent formation histories are realistic, at least for some galaxies
%

Another important issue in the solution of the ``angular momentum catastrophe" in our model is the fact that angular momentum is mainly lost from the smaller baryonic clumps. As reported in Sec. 3.1, the ``ordered" angular momentum, $h$, acquired by an halo is anticorrelated with the peak height $\nu$, and consequently smaller haloes acquire larger angular momentum. 
Similarly, the ``random" angular momentum $j$ is anticorrelated with $\nu$ (see Eq. C13, and Fig. 8 in DP09). Consequently, 
smaller baryonic clumps inside the halo have larger angular momentum (``random" angular momentum) than larger clumps. In the virialization process, these clumps loose more angular momentum than larger clumps, explaining why the low tail in the AMD is missing in Fig. 4.
}

%
%%the complex interplay of baryons and DM which acts similarly to the feedback mechanism in Maller \& Dekel (2002), and Governato %%et al. (2010). 
%

% TOLTO
%%The differences among the predicted distribution of angular momentum in our model and in N-body simulations has already been %%discussed in DP09 (Sect. 3), and also in Williams et al. (2004). As has already been written, the differences are due to the %%baryons DM interplay, not considered in purely N-body simulations, and are moreover connected to the different way haloes are %%obtained in N-body simulations and SIMs. In N-body simulations, haloes are obtained as a result of mergers of sub-haloes, while %%in our model haloes form as a result of a quiescent accretion of matter. The existence of many spiral galaxies with thin disks %%somehow supports the previous quiescent formation mechanism. Moreover, Dekel et al. (2009), showed that disc galaxies, at high %%$z$, are not formed through galactic mergers, but by smooth accretion flows.   
%

% TOLTO
%%In the same section of DP09, in Fig. 2, we showed that if we use our model to re-obtain a NFW profile, the angular momentum  %%distribution is similar to that coming out from N-body simulations, namely more centrally concentrated, and in disagreement with %%observations. The angular momentum distribution obtained for a typical halo coming out from our model has a different %%distribution, as previously described in this subsection. 
%

\subsection{The Too Big to Fail and the Missing Satellites problems}

The other small scale problem of the $\Lambda$CDM model is the MSP, concerning the incorrect
prediction of simulations of the subhaloes distribution in structures or satellites in our MW. 
The TBTF problem is a peculiar feature of the MSP, 
%concerning the mismatch between the central density of satellites in the simulations and observations.  
%USIAMO I LORO SATELLITI
which showed that subhalos in a DM-only simulation have significantly higher central densities than inferred from observations
of the Milky Way's classical dwarf spheroidal satellites.

Several solutions have been proposed to both problems (Strigari et al. 2007; Simon \& Geha 2007; Madau et al. 2008; Zolotov et al. 2012; Brooks \& Zolotov 2012; Purcell \& Zentner 2012; Vera-Ciro et al. 2012; Di Cintio et al. (2012); Wang et al. 2012). B13, proposed an interesting baryonic solution to the MSP and the TBTF problem.  
Namely, instead of running SPH simulations of different galaxies, they tried to introduce the baryons effect in large N-body dissipationless simulations, like the VL2, showing that the result obtained is in agreement with observations of MW and M31 satellites. 

{In the following, we will partly follow their steps to obtain the corrected circular velocities and distribution of VL2 satellites in the framework described in Sect. 2.

In summary the method is based on the following ideas and is divided into two main phases. 

Initially the attention is directed to the isolated satellites, before accretion to the main system. As noticed by several authors (e.g., Mashchenko et al 2006, 2008; Pe\~narrubia et al. 2010) the effects of the tidal forces of the main halo on a satellite depends fundamentally on its shape. If the satellite has a cuspy profile, its structure will not suffer big changes, when it will enter the main halo. If the satellite has a cored profile, the tidal field of the main halo can strip easily its gas and in some cases even destroy it (Pe\~narrubia et al. 2010). So, this first phase, defining the shape of the satellite is of fundamental importance. This first phase and the flattening of the satellites density profiles will be studied through our model. 

Then comes the second phase, when the satellite is no longer considered isolated, it is subject to the tidal field of the main halo, and finally it is accreted to the main halo (see the following). 

%
%In order to realize the previous plan, we will use our analytic approach of Sect. 2, and apply it to the satellites in Z12, in %order to test how core creation affects the missing satellites problem. Our approach gives to us the flattening of the profile %(core creation) before infall. 

Our result for the flattening of the density profile (first phase), is expressed in terms of the change to circular velocity in 1 kpc, as in Z12, and is given by 
%Z12-like correction is 
%Comparing the galaxies obtained with our model in presence of baryons with the same objects obtained in absence of baryons, we got a correction at $z=0$ %for the circular velocity $v_c$ at 1 kpc, $v_{1kpc}$
\begin{equation}
\Delta(v_{1kpc})= {\rm 0.3 v_{infall}-0.2 km/s }
\label{eq:mia}
\end{equation}
in the ${\rm 20 km/s<v_{infall}< 50 km/s }$ interval.
We want to stress again that the previous correction is determined for the Z12 satellites, and this correction will be applied to VL2, together with the other corrections discussed in the following. The correction that we obtained is close to that obtained by Z12, 
%Before discussing the second phase (changes in the satellite due to interaction with the main halo), we recall how Z12 obtained %the correction to the satellites shapes (hereafter Z12 correction), and compare their result with ours. 
%The correction by them carried to the VL2 satellites are: 1) 
%The Z12 correction, 
accounting mainly for the reduction of the subhaloes central mass produced by SF,
% and tidal stripping, 
%; b) gas loss produced by heating and stripping during the passage of the satellite in the main system; c) tidal stripping %enhancement due to interactions with the disc of the main system. 
and given by
\begin{eqnarray}
\Delta(v_{1kpc}) &=& {\rm 0.2 v_{infall}-0.26 km/s }
\nonumber \\
& & {\rm 20 km/s<v_{infall}< 50 km/s  }
%\Delta(v_{1kpc}= 0.2 v_{max,DM-only}-0.26 km/s
\end{eqnarray}

The last equation gives the difference between DM and SPH runs, and then the corrections to apply to satellites in N-body simulations that take account of the missing piece of baryonic physics. 
%Tolto
%including both SF and tidal stripping. 
%TOLTO
%%As previously reported, the Z12 correction is designed to be used by people who run a DM-only simulation and then want to %%understand how baryonic physics would have changed $v_{1kpc}$. It essentially allows to go from ``theory/simulations" to %%``observables".   
}

In the case of a $v_{infall} = {\rm 30 km/s }$, the Z12 correction gives $\Delta(v_{1kpc})=5.74$, while our correction gives $\Delta(v_{1kpc})=8.8$. The difference between the $\Delta(v_{1kpc})$ of our model and  
%{\bf CAUSA DIFFERENZA???? }
that of Z12 is due to the different models used to produce the pre-infall flattening of the satellites density profile. 
%TOLTO
%%As already reported, the Z12 correction is obtained taking into account the flattening of the profile before the infall, and %%adding to this the further flattening at the infall. 
The difference among our model and that used in Z12 is the model used to calculate the pre-infall flattening. In Z12 the pre-infall flattening is due to SF, and in our case is connected to dynamical friction. As shown by Cole et al. (2011), DF on infalling clumps is a very efficient mechanism in flattening the DM profile. A clump having a mass of 1\% of the halo mass can give rise to a core from a cuspy profile removing two times its mass from the inner part of the halo. In the case of the SF the mechanism should be less effective going down to lower masses (e.g., dwarfs with stellar mass $< 10^5-10^7 M_{\odot}$ have few stars and supernovae explosions are as a consequence less present).

As reported by B13, the baryonic corrections taken into account by the Z12 correction, and consequently ours, reduces the number of massive satellites that are expected in dissipationless N-body simulations, but we are also interested in seeing if the baryonic corrections also reduce the number of luminous satellites, and for this other corrections are needed. 

%Before going on, we would like to stress an important issue concerning the Z12 correction. 

Our correction (Eq. \ref{eq:mia}), similarly to the Z12 correction
%The last 
applies to satellites that survived at $z=0$, and that had the central $V_c$ reduced by baryonic physics.
%due to tidal stripping and SF.
Before $z=0$ satellites could have been destroyed (by e.g. stripping or photo-heating) (see the following). In N-body simulations, like the VL2, baryons effects are not taken into account, and then satellites that in the real universe or in SPH simulations may be totally destroyed by enhanced tidal stripping (due to disks and cores) and SF,
%SNe feedback, 
will not be destroyed in the VL2. Since our correction and Z12 correction applies to
%is only for 
satellites that survived to z=0, not those that were previously destroyed, we need two other corrections (tidal stripping, and suppression of star formation) 
%hence the need for the other two corrections is necessary, 
as done by B13, to calculate the survived satellites and then apply the Z12 correction.

{Here starts the second phase always mentioned, considering the effects of the interaction between the main halo and the satellite.

In order to take account of tidal stripping, and tidal heating after infall, to get accurate $V_{max}$ values, we could follow 
%two different approaches: a) to use 
%In the first one, we could use 
the Taylor \& Babul (2001) model (see the following), which takes into account the effect of tidal mass loss, and tidal heating in an analytical manner. 
However, this will be done in a future paper, and here we estimate the effect of stripping after infall, adding two corrections as done in B13.
%when the satellite enters the galaxy.

The first of the two corrections is the suppression of star formation by photo-heating, obtained following the Okamoto et al. (2008) results. The last one is the destruction rates by stripping of satellites, obtained following 
Pe\~narrubia et al. (2010).

We will use Eq. 8 from Pe\~narrubia et al. (2010), for subhaloes with density slope of dark matter $\gamma$, to calculate 
\begin{equation}
v_{max}(z=0)=v_{infall} \frac{2^\alpha x^\eta}{(1+x)^\alpha},
%v_{max}(z=0)=v_{infall} \frac{2^0.4 x^0.3}{(1+x)^0.3}
\end{equation}
where $x \equiv {\rm mass(z=0)/mass(z=infall) }$, describing the change in $V_{max}$ as a function of mass lost to tidal stripping, and $\alpha$, and $\eta$ are two fitting parameters whose values are connected to the slope $\gamma$ as shown in Fig. 6 of the Pe\~narrubia et al. (2010).     
For the DM-only runs, which have cuspy density profiles, $\gamma =- 1$ gives a good fit to the change in $V_{max}$. Since the satellites have a profile that can be cored, we have chosen a value of $\gamma$ in the range $\gamma=-0.5 \div 0$, and values of 
$\alpha$, and $\eta$ given by Pe\~narrubia et al. (2010), Fig. 6. In the case of the VL2 satellites, the values of $\alpha$ and $\eta$ are 0.4, and 0.3, respectively. 

%For the SPH runs, particularly the ones with lots of star formation that should be cored, $\gamma = 0$ is a good description.
In order to use the quoted equation, we need to know the mass lost in tidal stripping. Those values can be read from Fig. 6 in Z12. Then, we apply the 
Pe\~narrubia et al. (2010) equation, to get the change in $V_{max}$. Assuming that $V_{max}$ is approximately the same as $V_{1kpc}$, 
%(this is not a bad assumption), 
we get the needed correction after infall.

}

%{\bf corrected}
Similarly to B13, a) the population of satellites that looses more than 97\% of their infall mass, and b) all subhalos losing more than 90\% of their infall mass, having also apocentric passages ${\rm <20 kpc }$ and $v_{infall} > {\rm 30 km/s }$, are considered destroyed. Then the subhaloes that remain dark because they are affected by UV heating, and do not form stars, are calculated using the Okamoto et al. (2008) SPH simulations and Sect. 3.2 results of B13. Luminosity is assigned to the VL2 satellites as done in B13 (Eqs. 3, 4).

Finally, we have to assign a luminosity to the satellites which survived. We use the relation between $v_{\rm infall}$ and the stellar mass $M_\ast$ obtained by Z12
\begin{equation}
(\frac{v_{\rm infall}}{\rm kms^{-1}})^6=55.56  \frac{M_{\ast}}{M_{\odot}}
\end{equation}
and then using a relation between $M_{\ast}$ and the visual magnitude $M_{\rm V}$
\begin{equation}
M_{\ast}= L_{\rm V} \times 10^{-0.734+1.404(B-V)}
\end{equation}
where 
\begin{equation}
L_{\rm V}= 10^{-(V-4.8)/2.5}
\end{equation}
Munshi et al. (2013), that we approximate as in B13 with
\begin{equation}
\log_{10}(\frac{M_{\ast}}{M_{\odot}})=2.37-0.38 M_{\rm V}
\end{equation}

The result of the previously corrections are plotted in Fig. 6. The top panel of this figure represents the 
results from VL2 at $z=0$. 
The bottom panel represents the same satellites after the previous discussed corrections (heating, destruction, and velocity corrections) were applied. 
The red filled symbols are the object ``observable" in VL2. Dark objects are marked by empty circles. 
Objects marked with a circle and a ``x" through them, represent subhaloes that do not survive to the baryonic effects (e.g., baryonic disc, etc). 
Empty circles have a mass smaller than the minimum to retain baryon and form stars.
Finally, filled black circles are satellites that lose 90\% of their mass since infall, but do not satisfy the destruction criteria previously described.

Note that the Z12 correction was not applied to satellites with $v_{max}> {\rm 50km/s }$ (for example to the satellites with $M_{V}< -16$, that are the 
5 most massive satellites at infall, and are the five that had $v_{max}> {\rm 50km/s }$).

The number of satellites with $v_{1kpc}>20$ km/s is, similarly to B13, equal to 3.
%
%%, and more luminous than the more luminous MW's brightest dSph, namely Fornax TOGLIERE?. 
%
Differently from B13, our central velocities are smaller (the correction to the circular velocity, $\Delta(v_{1kpc})$, to subtract to the circular velocity itself, $V_c$, is larger in our model with respect to Z12, and B13). Also in our case some satellites are ``overcorrected", namely their velocities are negative. 

%{\bf a) WHY OUR CENTRAL VELOCITIES ARE SMALLER THAN B13; b) WHAT IS THE MEANING OF THE OVERCORRECTION }

The reason for the quoted overcorrection is similar to that described in B13. The halos whose velocity is overcorrected, is a halo population that after infall lost a great part of their mass and  so, at $z=0$, the circular velocity in 1 kpc, $v_{1kpc}$, is very low, giving rise to the quoted overcorrection. 
After infall, the quoted population lost more than 99.9\% of their mass, and moreover their tidal radius is $<1$ kpc. The population can be considered a population of destroyed subhaloes.  

%Regardless of this, we think they would have all been fully tidally disrupted in the presence of the disk tidal stripping.
%%\begin{figure}
%%\psfig{file=mom_ang3.eps,width=10.0cm}
%\psfig{file=ff.eps,width=15.0cm}
%\psfig{file=wi33.eps,width=8.0cm}
%\psfig{file=figamdr4.eps,width=8.0cm}
%%\caption[]{Distribution of the total specific angular momentum, $J_{Tot}$. The dotted-dashed and dashed line represents the %%quoted distribution for the halo n. 170 and n. 081, respectively, of van den Bosch et al. (2002). The dashed histogram is the %%distribution obtained from our model for the $10^{12} M_{\odot}$ halo and the solid one the angular momentum distribution for %%the density profile reproducing the NFW halo, as descried in Section 4.
%%}
%%\end{figure}

\begin{figure}
\psfig{file=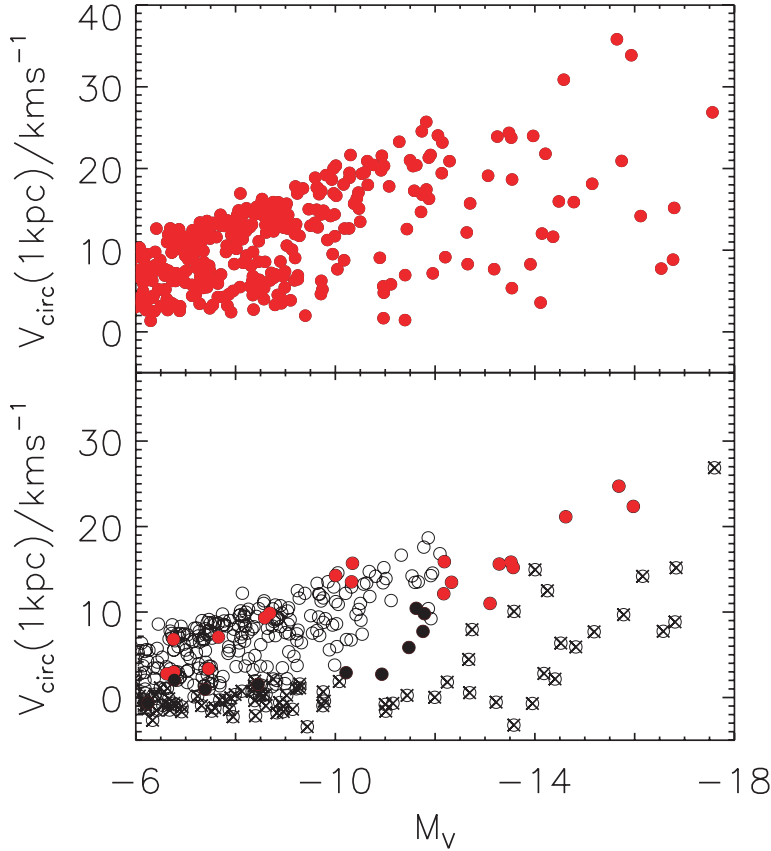,width=10.0cm}
\caption[]{Plot of $v_{1kpc}$ vs $M_V$ for the VL2 simulation subhaloes. In the top panel, we plot the VL2 satellites vs $M_V$ at $z=0$, as in B13. In the bottom panel, we represent the same VL2 satellites of the top panel with the baryonic corrections described in the text. The filled black circles represent satellites that have lost enough mass, stars are stripped and the luminosities are upper limits.
%{\bf stars should be stripped and the luminosities should be considered upper limits.}
Filled red circles are satellite observed at $z=0$, while circles with an x are subhaloes that have low probability to survive to tidal effects. Dark subhaloes are represented by empty circles. 
}
\end{figure}

As in B13, UV heating, and tidal destruction are necessary to reconcile the total number of luminous satellites with observations, while the Z12 correction is necessary to reconcile the masses of the subhaloes with observations. 

In the present section, we considered the evolution of satellites before infall, and calculated the flattening of the density profile using our model (not the SF flattening effect), and added to this the further flattening (or reduction of the central rotation velocity) produced by the interaction with the main halo. This last step was based on the calculation of the flattening, after infall, shown in B13. In a forthcoming paper, we will improve the model as follows. We calculate again the profile flattening of satellites before infall as done in this paper, but then we will use a model, like that of Taylor \& Babul (2001) able to follow the merger history and growth of the satellites while interacting with the main halo, tracking the substructure evolution in the main halo (Del Popolo \& Gambera 1997), taking account of the mass loss due to tides, and tidal heating. In this way, we have enough information to conclude if they are massive enough to retain baryons and form stars. 

Since our model is not so computationally ``heavy" as SPH simulations, we could study the MSP problem in different galaxies, since it is not enough to solve the problem in a single galaxy to conclude that the problem is solved in galaxies different from ours. In fact, several authors have discussed the MSP problem in relation to the host galaxy mass. Di Cintio et al. (2012), Vera-Ciro et al. (2012), Wang et al. (2012), showed that if the MW true virial mass is smaller than $10^{12} M_{\odot}$, namely $\simeq 8 \times 10^{11} M_{\odot}$, the satellites excess may disappear. 

In any case, the main point of our study, even if some points of the previous results need to be improved, is that baryonic physics is able to solve simultaneously the small scale problems of the $\Lambda$CDM model: cusp/core problem, angular momentum catastrophe, MSP, and TBTF problem. 

A similar result, has been obtained by a series of papers (Governato et al. 2010; Governato et al. 2012; Zolotov et al. 2012; Brooks \& Zolotov 2012). In those papers, 
%the main effect producing the flattening of the cusp, and the MSP solution is connected to the SF effect. 
the main effect producing the flattening of the cusp, and the TBTF solution is connected to the effects of SF, 
%SNe feedback, 
following episodic bursts of supernovae.

In our model, the solution to the previous quoted problems is connected to the complex interaction between DM and baryons mediated by DF. Our study is similar to those of El-Zant et al. (2001, 2004), Romano-Diaz et al. (2008), Cole et al. (2011), in the sense that like them in our study DF plays an important role. 
{However, differently from the previous studies, we consider the joint effect of several other effects (e.g., random angular momentum, angular momentum generated by tidal torques, adiabatic contraction, cooling, star formation),  that in the quoted studies were not considered. }

{Before concluding this section, we want to recall that recently Ferrero et al. (2012) showed that the field dwarf galaxies have a similar behavior as the satellites. In order to reconcile the fact that the galaxy stellar mass function is less steep than that of dark matter halos, it is necessary to assume that the efficiency of galaxy formation decreases in a sharp fashion with decreasing halo mass. This implies that there should be a clear evidence that dwarfs having noteworthy differences in stellar mass are hosted in haloes spanning a narrow range in mass. This is not observed in their study, and moreover a large part of the galaxies by them studied have much less dark mass enclosed in the dwarf.
%%%%%dark mass much lower than what is usually expected from halos as massive as $10^{10} M_{\odot}$. 
One of the possibility to solve the ``puzzle" is that baryonic physics, like the blow-outs produced by SF, previously discussed. However, especially in the smallest systems this solution seems very unappealing, since it is difficult to explain how baryons in a galaxy with stellar mass similar to that of globular clusters can change the inner structure of a $10^{10} M_{\odot}$ halo. 
%In particular they showed that there is no evidence that dwarfs having noteworthy differences in stellar mass are hosted in haloes with mass 

This objection moved from Ferrero et al. (2012) to the SF model, was already discussed in Sec. 3.3 of this paper. In that section, we reported that one has to expect that the SF mechanism becomes less effective in smaller galaxies, since in dwarfs having 
stellar mass $< 10^5-10^7 M_{\odot}$ a supernova explosion is less probable than in a larger galaxy having a larger stellar mass. 

The problem envisioned by Ferrero et al. (2012) for the SF mechanism 
%last problem 
is less important in our model since the dark matter mass reduction process obtained through our model is more efficient than that connected to SF. As reported in Sec. 3.3, clumps of 1\% of the halo mass can remove two times their mass, flattening cuspy profiles. 
}

\section{Conclusions}

In the present paper, we studied the small scale problems of the $\Lambda$CDM by means of the model presented in DP09 (see also DP12a, b) following the proto-structure evolution with a semi-analytical model taking into account the effect of adiabatic contraction, dynamical friction and the exchange of angular momentum between baryons and DM, ordered and random angular momentum.

The model had already shown in DP09, DP12a,b, that the angular momentum got by the system through tidal torques and random velocities (random angular momentum), can be transferred in part to the DM from 
{baryonic clumps of 1\% the mass of the halo} through DF action. This produces a flattening of the cusp in agreement with previous studies based on DF driven flattening of the cusp (El-Zant et al. 2001, 2004; Romano-Diaz et al. 2008) and SF driven flattening of the cusp (Navarro et al. 1996a; Gelato \& Sommersen-Larsen 1999; Read \& Gilmore 2005; Mashchenko et al. 2006, 2008).

{The mass of the quoted clumps, similar to those used by El-Zant et al. (2001), and Cole et al. (2001) is quite large. In the case of the MW, using data from McMillan (2011), 1\% of the halo mass of the MW, $1.26 \pm 0.24 \times 10^{12} M_{\odot}$, corresponds to $\simeq 10^{10} M_{\odot}$ close to the total star content ($6.43 \pm 0.63 \times 10^{10} M_{\odot}$) or the disc mass ($\simeq 2.4 \times 10^{10} M_{\odot}$). We want to recall, that nevertheless the previous assumption, lighter clumps are more efficient. As shown by Cole et al. (2011) in their Fig. 1, the ratio among the maximum ``excavated" mass to the mass of the clump increases with decreasing clump mass. While a clump of 1\% of that of the halo can excavate $\simeq 2$ times its own mass, a clump with mass 0.001\% of that of the halo can excavate till $\simeq 5$ times its own mass depending on the central density slope and the structure scale radius (see Eq. 11 of Cole et al. 2011). }
~\\

{Another assumption, used in the paper is that haloes form primarily through quiescent accretion, in agreement with the prescriptions of spherical collapse model. As previously discussed the existence of thin bulgeless disks support the previous idea, but at the same time we know that the role of mergers, in general, is important in the overall formation of structures. 
~\\

We want to stress that our results are strictly depending on the previous two ad hoc assumptions.
}
~\\

The MSP and TBTF problem were discussed in the same framework as follows. We studied through our model how baryons influence the inner density profile (or the inner circular velocity, $v_{1kpc}$) before infall in the host halo. Then, we added the changes to 
$v_{1kpc}$ after infall, produced by tidal stripping, and UV heating. We applied these corrections to the VL2 satellites, obtaining the result  
%with the result 
that their mass and number is reconciled with observations. 

% VEDERE SE RIMETTERLO
%%We also checked if SF had a fundamental role in .......

We also compared the distribution of spin, and the angular momentum distribution, obtained from our model, to those obtained for 14 dwarfs by VBS finding that the DM and baryons distribution is different and we did not observe the lost of angular momentum typical of many past SPH simulations.

\section*{Acknowledgements}

We would like to thank the International Institute of Physics in Natal for the facilities and hospitality, Adi Zolotov, A. M. Brooks, and Charles Downing (from Exeter University) for helpful comments and for a critical reading of the paper. DCR and JCF thank CNPq and FAPES for partial financial support.

%\begin{thebibliography}{999}

\end{document}